\documentclass[apj]{emulateapj}

\begin{document}

\shortauthors{Gordon et al.}
\shorttitle{MIPS 70~\micron\ Calibration}	
\slugcomment{PASP, in press}

\title{Absolute Calibration and Characterization of the \\ Multiband
Imaging Photometer for Spitzer. II. 70 micron Imaging} 

\author{Karl~D.~Gordon\altaffilmark{1}, 
   Charles~W.~Engelbracht\altaffilmark{1},
   Dario~Fadda\altaffilmark{2},
   John~Stansberry\altaffilmark{1},
   Stefanie~Wachter\altaffilmark{2},
   Dave~T.~Frayer\altaffilmark{2},
   George~Rieke\altaffilmark{1},
   Alberto~Noriega-Crespo\altaffilmark{2},
   William~B.~Latter\altaffilmark{3},
   Erick~Young\altaffilmark{1},
   Gerry~Neugebauer\altaffilmark{1},
   Zoltan~Balog\altaffilmark{1},
   Herv\'e~Dole\altaffilmark{6}, 
   Eiichi~Egami\altaffilmark{1},
   Dean~Hines\altaffilmark{4},
   Doug~Kelly\altaffilmark{1},
   Francine~Marleau\altaffilmark{2},
   Karl~Misselt\altaffilmark{1},
   Jane~Morrison\altaffilmark{1},
   Pablo~P\'erez-Gonz\'alez\altaffilmark{1,5},
   Jeonghee~Rho\altaffilmark{2}, and
   Wm.~A.~Wheaton\altaffilmark{2}
   }
\altaffiltext{1}{Steward Observatory, University of Arizona, Tucson, AZ 85721}
\altaffiltext{2}{Spitzer Science Center, 220-6, Caltech, Pasadena, CA 91125}
\altaffiltext{3}{NASA Hershel Science Center, 100-22, Caltech, Pasadena, CA 91125}
\altaffiltext{4}{Space Science Institute, 4750 Walnut Street, Boulder, CO 80301}
\altaffiltext{5}{Departamento de Astrof\'{\i}sica, Facultad de CC. 
F\'{\i}sicas, Universidad Complutense de Madrid, E-28040 Madrid, Spain}
\altaffiltext{6}{Institut d'Astrophysique Spatiale (IAS), b\^at. 121,
91405 Orsay Cedex, France} 

\begin{abstract} 
The absolute calibration and characterization of the Multiband Imaging
Photometer for Spitzer (MIPS) 70~\micron\ coarse- and fine-scale
imaging modes are presented based on over 2.5 years of observations.
Accurate photometry (especially for faint sources) requires two simple
processing steps beyond the standard data reduction to remove
long-term detector transients.  Point spread function (PSF) fitting
photometry is found to give more accurate flux densities than aperture
photometry.  Based on the PSF fitting photometry, the calibration
factor shows no strong trend with flux density, background, spectral
type, exposure time, or time since anneals.  The coarse-scale
calibration sample includes observations of stars with flux densities
from 22~mJy to 17~Jy, on backgrounds from 4 to 26~MJy~sr$^{-1}$, and
with spectral types from B to M.  The coarse-scale calibration is $702
\pm 35$~MJy~sr$^{-1}$~MIPS70$^{-1}$ (5\% uncertainty) and is
based on measurements of 66 stars.  The instrumental units of the MIPS
70~\micron\ coarse- and fine-scale imaging modes are called MIPS70 and
MIPS70F, respectively.  The photometric repeatability is
calculated to be 4.5\% from two stars measured during every MIPS
campaign and includes variations on all time scales probed.  The
preliminary fine-scale calibration factor is $2894 \pm
294$~MJy~sr$^{-1}$~MIPS70F$^{-1}$ (10\% uncertainty) based on 10
stars.  The uncertainty in the coarse- and fine-scale calibration
factors are dominated by the 4.5\% photometric repeatability and the
small sample size, respectively.  The 5$\sigma$, 500~s sensitivity of
the coarse-scale observations is 6--8~mJy.  This work shows that the
MIPS 70~\micron\ array produces accurate, well calibrated photometry
and validates the MIPS 70~\micron\ operating strategy, especially
the use of frequent stimulator flashes to track the changing
responsivities of the Ge:Ga detectors.
\end{abstract}


\section{Introduction}
\label{sec_intro}

The Multiband Imaging Photometer for Spitzer
\citep[MIPS,][]{Rieke04MIPS} is the far-infrared imager on the Spitzer
Space Telescope \citep[Spitzer,][]{Werner04Spitzer}.  MIPS images the
sky in bands at 24, 70, and 160~\micron.  The absolute calibration of
the MIPS bands is complicated by the challenging nature of removing
the instrumental signatures of the MIPS detectors as well as
predicting the flux densities of calibration sources accurately at
far-infrared wavelengths.  This paper describes the calibration and
characterization of the 70~\micron\ band.  Companion papers provide
the transfer of previous absolute calibrations to the MIPS 24~\micron\
band \citep{Rieke07absir}, the 24~\micron\ band calibration and
characterization \citep{Engelbracht07MIPS24}, and 160~\micron\ band
calibration and characterization \citep{Stansberry07MIPS160}.
\citet{Engelbracht07MIPS24} also presents the MIPS stellar calibrator
sample that is used for this paper.  The calibration factors derived
in these papers represent the official MIPS calibration and are
due to the combined efforts of the MIPS Instrument Team (at Univ.\ of
Arizona) and the MIPS Instrument Support Team (at the Spitzer Science
Center).

The characterization and calibration of the MIPS 70~\micron\ band is
based on stellar photospheres.  The repeatability of 70~\micron\
photometry is measured from observations of two stars, at least
one of which is observed in every MIPS campaign.  The absolute
calibration of the 70~\micron\ band is based on a large network of
stars observed in the standard coarse-scale photometry mode with a
range of predicted flux densities and backgrounds.  In addition to the
coarse-scale observations, a small number of stars were observed in
the fine-scale photometry mode to allow the coarse-scale calibration
to be transfered to this mode.  The observations used in this paper
include both In Orbit Checkout (IOC, MIPS campaigns R, V, X1, and W)
and regular science operations (MIPS campaigns 1-29) with a cutoff
date of 3 Mar 2006.

The goal of the calibration is to transform measurements in
instrumental units to instrument independent physical units.  The goal
of the characterization is to determine if the calibration depends on
how the data are taken (e.g., exposure time, time since anneal) or
characteristics of the sources being measured (e.g., flux density,
background).  The primary challenge for the 70~\micron\
characterization and calibration is accurately correcting the detector
transients associated with Ge:Ga detectors.  The standard reduction
steps are detailed in \citet{Gordon05DAT} but extra steps to achieve
accurate photometry with the highest possible signal-to-noise for
point sources are needed and discussed in this paper.

Accurate calibration and high sensitivity is important at 70~\micron\
as they enable a number of science investigations including the
detection and study of cold disks around stars \citep[e.g.,][]{Kim05,
Bryden06}, imaging of warm dust in galaxies
\citep[e.g.,][]{Calzetti05, Dale05, Gordon06}, and investigation of
faint, redshifted galaxies
\citep[e.g.,][]{Dole04a, Frayer06}.

\section{Data}
\label{sec_data}

The 70~\micron\ calibration program is based on stars with spectral
types from B to M, predicted flux densities from 22~mJy to 17~Jy, and
predicted backgrounds from 4 to 26 MJy/sr \citep{Engelbracht07MIPS24}.
The observations were carried out in photometry mode with 3.15 to
10.49~s individual image exposure times and a range of total exposure
times from $\sim$50~s to $\sim$560~s.  The majority of the
observations were performed in standard coarse-scale photometry mode
with a few done in the standard fine-scale photometry mode.  The
coarse- and fine-scale photometry 
modes are also referred to as the wide- and narrow-field photometry
modes.  

The coarse-scale mode samples the 18\arcsec\ FWHM 70~\micron\ PSF with
9.85\arcsec\ pixels.  The minimum coarse-scale photometry mode
observation consists of 12 images of the target and 4 images where the
internal calibration stimulator is flashed \citep[see Fig. 2
of][]{Gordon05DAT}.  The target point source is dithered around the
central part of good half of the 70~\micron\ array so that the source
is on different pixels for each 
of the 12 image exposures.  The stimulator flashes are used to remove
the responsivity variations in the Ge:Ga detectors by dividing each
image exposure by an interpolated stimulator flash.  This division
converts the raw DN/s units to fractions of the
stimulator flash amplitude (also measured in DN/s units) which are
termed MIPS70 units.  These MIPS70 units are surface brightness units
as the varying pixel size across the array has been normalized out
due to the division by the stimulator flash.  Readers should refer to
\citet{Rieke04MIPS} and 
\citet{Gordon05DAT} for the details on how MIPS data are taken and
reduced.  The maximum image exposure time is 10.49~s, thus longer
total exposures on a source are acquired by repeating the minimum set
of images described above.

The fine-scale mode samples the same 70~\micron\ PSF with
5.24~\arcsec\ pixels and is designed for detailed studies of source
structure.  The dithering strategy is different for fine-scale mode
where source-background pairs of images are acquired instead of
dithering the source around the array.  The minimum fine-scale
photometry mode observation consists of 8 source-background pairs of
images, 4 stimulator images, and 2 dedicated stimulator background
images.  The data reduction is the same as the coarse-scale mode and,
thus, the resulting raw units of the images are also fractions of the
stimulator flash and are termed MIPS70F units.  These units are
different than the MIPS70 units due to the change in the optical train
used.

The coarse-scale observations were extensive and motivated to check
non-linearities versus flux density, background, exposure time, etc.
The fine-scale observations were done to transfer the coarse-scale
calibration to the fine-scale.  The coarse-scale and scan map modes share the
same optical train and only differ in the dithering strategy, thus the
coarse-scale photometry calibration should apply to the scan map mode
observations.

\subsection{Data Reduction}

\begin{figure*}
\epsscale{1.1}
\plotone{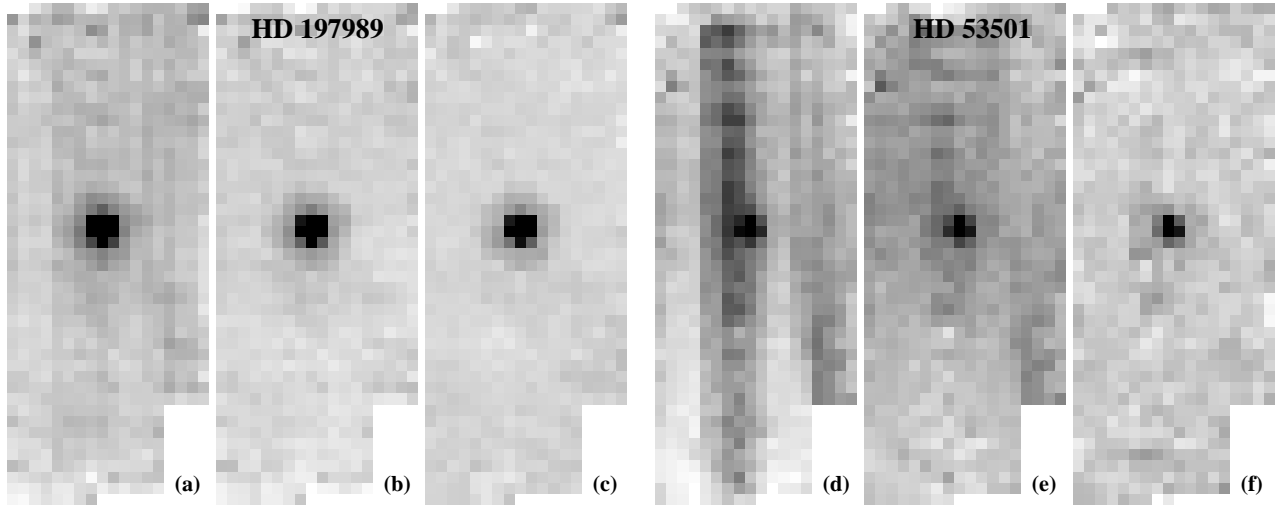}
\caption{The mosaics of coarse-scale observations of two stars of different
brightness are shown with (a,d) default reductions, (b,e) column mean
subtraction, and (c,f) column mean subtraction and time filtering.  The
two stars are HD~197989 (AOR \#13590784, flux density = 787~mJy) and HD~53501
(AOR \#13641984, flux density = 135~mJy).  The HD~197989 images are displayed
with a linear stretch that ranges from 0.022 to 0.05 (a), -0.005 to
0.025 (b), and -0.005 to 0.0025 (c).  The HD~53501 images are displayed
with a linear stretch that ranges from 0.009 to 0.025 (d), -0.005 to
0.01 (e), and -0.002 to 0.008 (f).  The images and ranges are all given in MIPS70
units.
\label{fig_red_example} }
\end{figure*}

Each observation was reduced through the MIPS Data Analysis Tool
\citep[DAT, v3.06,][]{Gordon05DAT}.  The resulting mosaics of this default
processing are shown in Fig.~\ref{fig_red_example}a,d for two point
sources observed in coarse-scale mode.  It is possible to improve the
detection of point sources taken in photometry mode by utilizing the
redundancy of the observations to remove residual instrumental
signatures.  These residual signatures arise because the stimulator
flashes calibrate the fast response of the detectors well, but there
is a drift between the fast and slow response of the detectors
\citep{Haegel01, Gordon05DAT}.  In coarse-scale mode, point sources
are dithered around the array to ensure their signals are in the well
calibrated fast response.  The dithering does not put the background
signal in the fast response and as the sky level is roughly equal at
the different dither positions, the background is in the slow response
regime.  As an accurate measurement of the background is essential for
good photometry, the drifting background needs to be corrected by two
additional steps.  The extra steps are designed not to introduce
biases into the data 
based on source flux density while reducing the residual instrumental
signatures.

The largest portion of the drift is seen to be in common among
pixels in the same column.  This is not surprising as columns
represent a common strip of detector material \citep{Young98}.  The
column offset can be removed easily for observations of isolated point
sources by subtracting the median of the column.  The median of each
column is computed after excluding a region centered on the source
with a diameter of 9 pixels ($\sim$89\arcsec) to ensure the resulting
correction is not biased by the source.  The column
subtracted data are shown in Fig.~\ref{fig_red_example}b,e.  A
smaller, but still significant background drift seen as array
dependent structure is still present after this correction.  This
smaller residual instrument signature can be removed from each pixel
by using a simple time filter.  This time filter works by subtracting
the mean value from a pixel of the previous and next 14 measurements
of that pixel.  To ensure that this mean value is not biased by the
presence of the source itself, all measurements of the source within a
spatial radius of 4.5 pixels as well the current, previous, and next
measurements are not used in computing the mean value.  The mean is
computed after excluding all the pixels $>4\sigma$ from the median (sigma
clipping).  The time filter only uses data taken on a particular
source which means that the time filtering is less accurate at the
beginning and end of the observation set.  This time filtering with
constraints was optimized to  
minimize the background noise.  The result is shown in
Fig.~\ref{fig_red_example}c,f.  From the two examples shown in
Fig.~\ref{fig_red_example}, it is clear that the importance of these
extra processing steps increases with decreasing source flux density.

Unlike coarse-scale photometry mode where the large field of view
provides sufficient area for background estimates, the reduced field
of view of the find-scale mode requires that the point source be chopped off the
array.  This simplifies the extra processing to just
subtracting the images taken in the off positions from the on position
images which effectively removes the column offsets and smaller scale
pixel dependent drifts in the background.

\subsection{Aperture Corrections}

\begin{figure*}
\epsscale{1.1}
\plotone{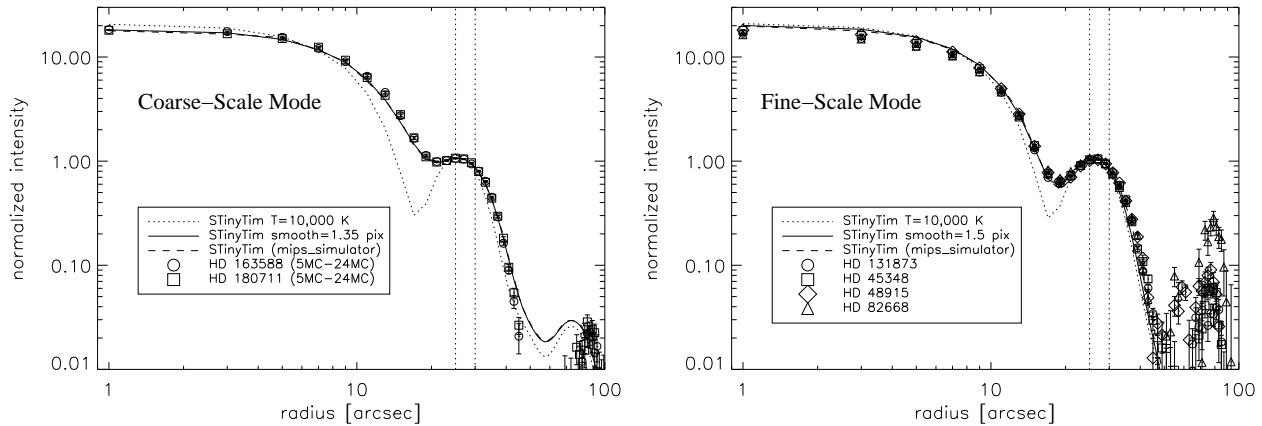}
\caption{The radial profiles of the observed and model PSFs are
plotted for coarse-scale and fine-scale modes.  The uncertainties on
the observed PSFs are calculated from the standard deviation of the
mean of the measurements in each measurement annulus.  The
PSF as predicted from STinyTim is shown as well as two smoothed model
PSFs.  All the PSFs have been normalized to one between radii of 25-30\arcsec\
(between the vertical dotted lines) with the background determined
from 100-200\arcsec\ for the coarse-scale and 70-100\arcsec\ for the
fine-scale.  In both plots, it is clear that the observed PSFs are
well measured out to around 40\arcsec\ where the PSF is about
1/1000 the brightness of the peak. 
\label{fig_psf_comp} }
\end{figure*}

A necessary part of measuring the flux density of a point source using
aperture photometry or point spread function (PSF) fitting is an
accurate PSF.  The repeatability measurements on HD~163588 and
HD~180711 provide the ideal opportunity to compare the STinyTim
\citep{Krist02} model of the MIPS 70~\micron\ PSF to the coarse-scale
observations.  The repeated observations of these two stars allow for
very high signal-to-noise observed PSFs to be constructed.  All of the
observations of these two stars taken after the final
optimization of the array parameters
($\geq$5th MIPS campaign) were mosaiced to produce two
empirical PSFs.  The fine-scale observed PSFs are from observations of
the fine-scale calibration stars.  The observed PSFs are compared with
the STinyTim PSF for a T~=~10000~K blackbody in
Fig.~\ref{fig_psf_comp}.  All stars in our calibration program have
the same spectrum across the 70~\micron\ band as the Rayleigh-Jeans
tail of stellar spectra is being sampled.  Thus, the PSF generated
assuming a T~=~10000~K blackbody is a good representation of any
star's PSF as long as it does not have an infrared excess.

As can be seen for both coarse- and fine-scale, the model PSF well
represents the observed PSFs when the smoothing associated with the
pixel sampling is applied.  We have simulated the pixel sampling
smoothing two different methods.  The first method used uses the
mips\_simulator program which produces simulated MIPS observations
with the observed dithering using an input PSF.  These simulated
observations where mosaiced, using the same software which is used for
the actual observations, to produce the mips\_simulator PSF shown.
The second method uses a simple boxcar smoothing function.  The
``smooth=1.35 pix'' and ``smooth=1.5 pix'' PSFs are created by
directly smoothing the STinyTim PSF with a square kernel with the
specified width where the pixel sizes are 9.85$\arcsec$ and
5.24$\arcsec$ for the coarse- and fine-scales, respectively.  The
close correspondence between the observed, mips\_simulator, and
directly smoothed PSFs means that accurate MIPS 70~\micron\ PSFs can
be generated from the STinyTim PSF smoothed with a simple square
kernel.

The correspondence between the observed and STinyTim model PSFs was
expected as the MIPS 70~\micron\ band should be purely diffraction
limited.  The 70~\micron\ band has a bandwidth of $\sim$19~\micron\
resulting in the PSF varying significantly between blue (peaking at
wavelengths shorter than the band) and red (peaking at wavelengths
longer than the band) sources.  In addition to stellar sources,
we have verified that STinyTim model PSFs describe the observed PSFs
for sources with blackbody temperatures as low as 60~K using
observations of asteroids and Pluto.  Having a valid model for the
PSF 
allows for accurate, noiseless aperture corrections as the total flux
density is known from PSFs created with different source spectra.

The observed fine-scale PSFs do show disagreement in the core, where
the model PSF is systematically higher.  This is not surprising given
that there are known flux density nonlinearities in the Ge:Ga
detectors that are not corrected in the standard data reduction and
the fine-scale calibration stars are brighter than those observed in
the coarse-scale mode.
Characterization of these flux density nonlinearities is ongoing, but
preliminary indications are that they are $\sim$15\% for point source flux
densities of $\sim$20~Jy observed in the coarse-scale mode (see
\S\ref{sec_nonlin}).  The coarse-scale 
observed PSFs do not show any systematic disagreement in the core,
consistent with the use of fainter stars.

The aperture corrections were calculated by performing aperture
photometry on square kernel smoothed model PSFs.  The model PSFs were
computed for a $64\arcmin \times 64\arcmin$ field to ensure that the
total flux of a point source was measured.  The method for computing
the aperture corrections is the same as used for measuring the
calibration star flux densities used in this paper.  The method does
not account for partial pixels but given that the model PSFs were
computed with pixels 10$\times$ smaller than the array pixel size this
is not expected to be an issue even for the smallest apertures
considered in this paper.  For the purposes of this paper, we use the
radius = 35\arcsec\ aperture (1.22 aperture correction) to minimize
the sensitivity of the 
calibration to uncertainty in the aperture correction and centering
errors.  The aperture corrections for a small sample of object
aperture and background annuli are given in Table~\ref{tab_apcor} for
three PSFs (T~=~10000~K, 60~K, and 10~K blackbodies).
Three PSFs with different source spectra are given to emphasize the
importance of using PSFs with the right source spectrum for accurate
photometry.

\subsection{Measurements}

The photometry was measured with aperture photometry and PSF fitting
for each observation of a calibration star.  The aperture photometry
was done with a circular aperture with a radius of $35\arcsec$ and a
sky annulus of 39 to $65\arcsec$.  The PSF fitting was done with
StarFinder \citep{Diolaiti00} which is ideally suited for the well
sampled and stable MIPS PSFs.  The flux densities measured with these
two methods are listed in Tables~\ref{tab_meas_wf}-\ref{tab_meas_nf}.
The aperture flux densities have had the aperture correction applied
and the PSF flux densities are naturally for an infinite aperture.  In
addition to the reported flux densities, signal-to-noise (S/N)
calculations are also reported.  In the case of the aperture
photometry, the S/N calculation is done using the noise in the sky
annulus to determine both the uncertainty due to summing the object
flux density as well as subtracting the background.  This S/N does not
include the contribution from the photon noise of the source as the
gain of Ge:Ga detectors is not a well defined quantity.  In the case
of the PSF photometry, the S/N calculation is done by the StarFinder
program utilizing the empirical uncertainty image calculated from the
repeated measurements of each point in the mosaic.  Only measurements
with S/N greater than 5 are reported in
Tables~\ref{tab_meas_wf}-\ref{tab_meas_nf}.  The columns in these
tables give the star name, campaign of observation, AOR \# (unique
Spitzer observation identifier), exposure start time, individual
exposure time, total exposure time, aperture flux density,
signal-to-noise, PSF flux density, and signal-to-noise.

The measurements used in this paper were all reduced with the DAT and
custom software based on the DAT results.  Comparisons were
done with measurements using data reduced with the Spitzer Science
Center (SSC) pipeline and similar software available from the contributed
software portion of the Spitzer
website\footnote{http://ssc.spitzer.caltech.edu/archanaly/contributed/}.
Note that the SSC pipeline reduced images are given in MJy sr$^{-1}$
and must be divided by the applied flux conversion factor (given by
the FITS header keyword FLUXCONV) to recover the instrumental MIPS70
or MIPS70F units.
The result of this comparison is that the two methods produce
equivalent results with a mean ratio of DAT to SSC aperture photometry
for the full sample
of $0.994 \pm 0.037$.

\subsection{Flux Density Predictions}

The predicted flux densities at 70~\micron\ were derived from the
24~\micron\ predictions presented by \citet{Engelbracht07MIPS24} using
24/70~\micron\ colors derived from models.  For each star, the ratio
of the flux densities at the effective filter wavelengths of
23.675~\micron\ and 71.42~\micron\ was computed for the appropriate
\citet{Kurucz93} model (using a power-law interpolation) and a
blackbody at the effective temperature of the star.  The model ratio
was taken to be the average of these two values, which typically
differed by 1-2\%, and the uncertainty was taken to be the difference
between them.  The predicted flux densities at 24~\micron\ were divided
by this ratio to compute the 70~\micron\ flux densities.  The
uncertainties on the flux density predictions were calculated by
adding in quadrature the uncertainties in the 24~\micron\ flux density
and in the 24/70~\micron\ color prediction.  The average predicted
backgrounds for 
the observations were estimated from Spitzer Planning Observations
Tool (SPOT) with the uncertainties giving the range when each target is
visible.  The full sample of calibration stars covers a wide range of
ecliptic and Galactic latitudes providing a large range in
backgrounds, but for any particular star the backgrounds vary by
$<$30\% for different dates of observation.
The flux density predictions and background estimates are
listed in Tables~\ref{tab_ave_wf} \& \ref{tab_ave_nf}.  These tables
also give each star's spectral type, the average of the measured
aperture and PSF flux densities and their associated signal-to-noises.
In addition, the average calibration factor determined using the
aperture and PSF fitting measurements is given (see
\S\ref{sec_nonlin}).

The zero point of the 70~\micron\ band at the effective filter
wavelength of 71.42~\micron\ is $0.778 \pm 0.012$~Jy in the
\citet{Rieke07absir} system.  It is important to note that the
70~\micron\ calibration is based on stars (10,000~K blackbody).  Thus,
measuring accurate 70~\micron\ flux densities for objects with
different spectral energy distributions requires the use of the color
corrections given in
\citet{Stansberry07MIPS160}.

\section{Results}
\label{sec_results}

\begin{figure*}
\epsscale{1.1}
\plotone{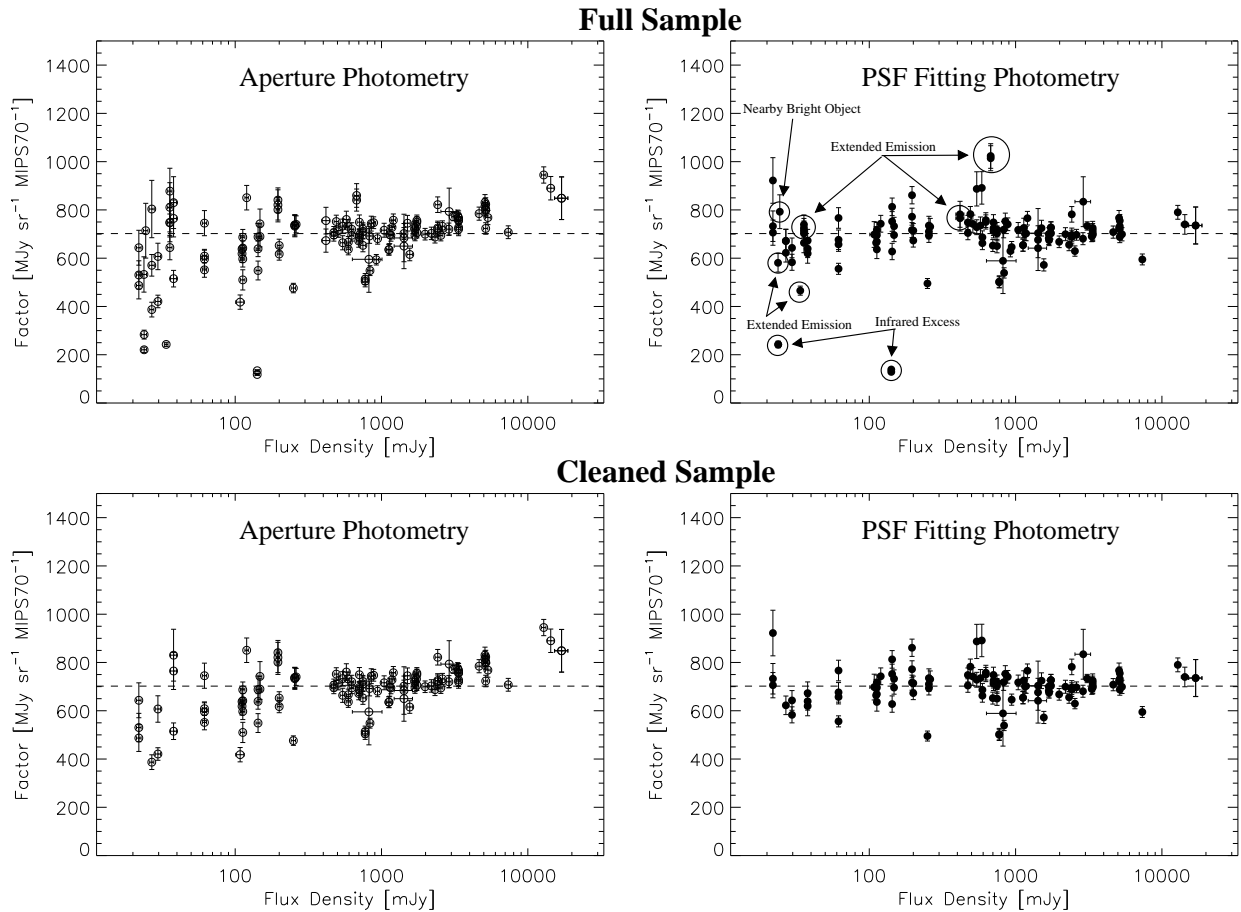}
\caption{The calibration factor for all the stars with positive flux
densities (expect for the two repeatability stars) are plotted versus
predicted flux density.  Each point gives the calibration factor for a
single measurement and, thus, a star can have multiple points.  The
uncertainties on each point include the measurement (y) and flux
density prediction (x \& y) uncertainties.  The dashed line is drawn
at the final calibration factor.  The cleaned sample is the same as the
full sample after the observations which have been rejected are
removed.  The reasons for rejecting a point are annotated in the
second panel of the full sample and
discussed in $\S$\ref{sec_results}.
\label{fig_calfac_vs_flux} }
\end{figure*}

The first step in deriving the final calibration factor is to examine
the ensemble of calibration factor measurements for stars which show
significant deviations.  These deviations can be the result of a star
having an infrared excess, a poor flux density prediction, or a
measurement corrupted by extended emission.  Infrared excesses and
poor flux density predictions can be seen by calculating the
calibration factor for each star and identifying stars that give
calibration factors that are significantly deviant from the ensemble.
Stars with excesses will have measured flux densities well in excess of the
predicted flux densities and, thus, will predict low calibration
factors compared to non-excess stars.  Figure~\ref{fig_calfac_vs_flux}
displays the calibration factors for all the calibration stars versus
predicted flux density.  The calibration factors are calculated as a
conversion from MIPS70 units to MJy~sr$^{-1}$ as the basic measurement
made by the 70~\micron\ detectors is flux per pixel which is a
surface brightness (see \S\ref{sec_calfac_wf} for the conversion from
MIPS70 units directly to flux densities assuming the default
70~\micron\ mosaic pixel size).  Measurements using both aperture photometry
and PSF fitting are shown in this figure.  The PSF fitting values show
less scatter than the aperture photometry values and, thus, will be
the basis for identifying deviant measurements.  We identify two stars
(HD~102647 and HD~173398) which have possible excesses from
Fig.~\ref{fig_calfac_vs_flux} and no stars with poor flux density
predictions.  There are seven stars that have significant extended
emission underlying or near the object position determined by visual
inspection of the 70~\micron\ images.  During this inspection, it was
seen that the observation of HD~173511 was affected by a nearby, much
brighter object and, as a result, this star was also rejected.  The
stars in our sample that have been rejected for any of these three
reasons are listed in Table~\ref{tab_reject}.  If a star was rejected
as part of the 24~\micron\ analysis given in
\citet{Engelbracht07MIPS24}, it is not used at all in this paper.

\subsection{Flux Density/Background Nonlinearities}
\label{sec_nonlin}

\begin{figure*}
\epsscale{1.1}
\plotone{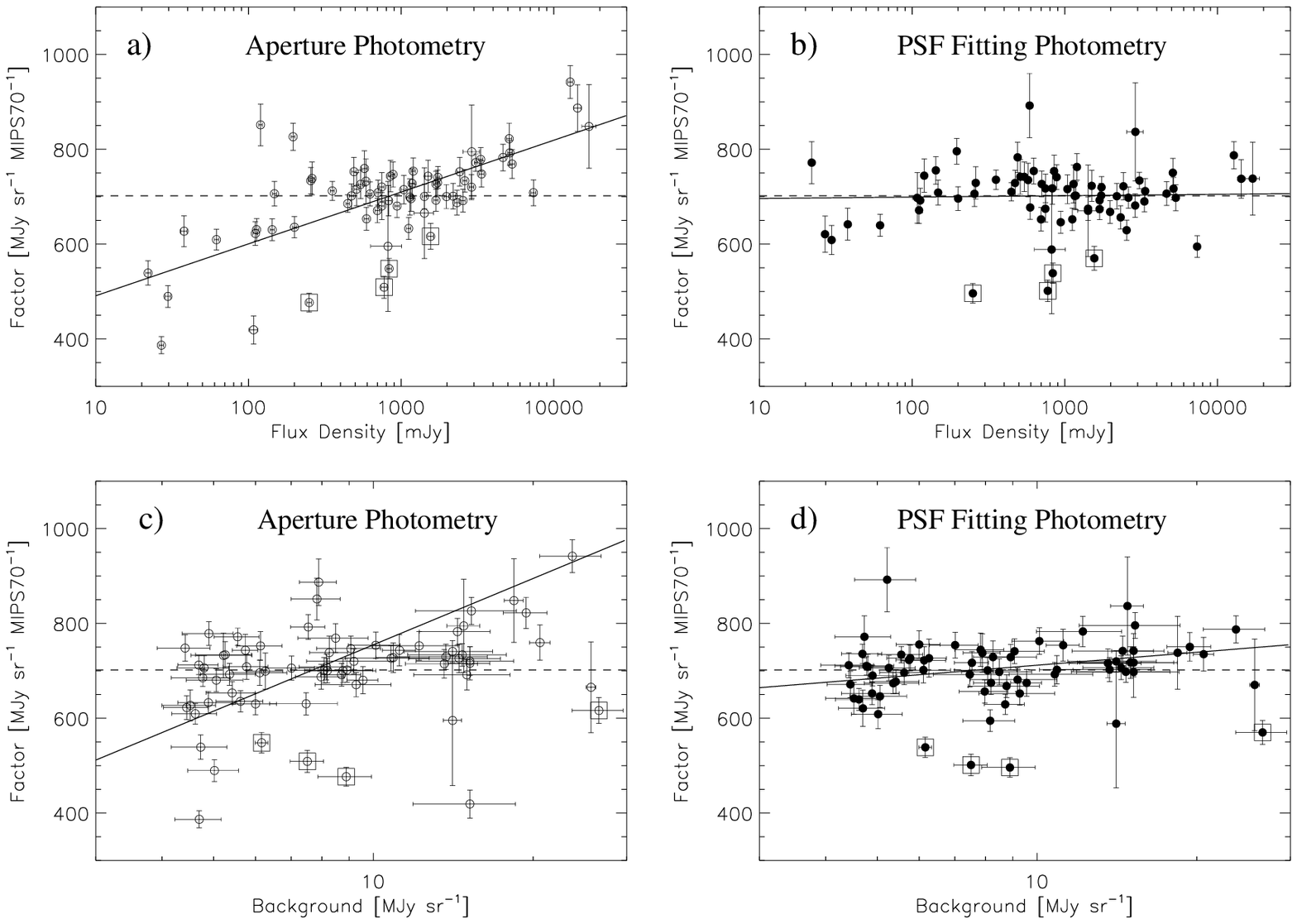}
\caption{The average calibration factor for each star is plotted
versus predicted flux density (a \& b) and background (c \& d).  The
dashed line is 
drawn at the final calibration factor.  The solid line gives the linear
fit to all the data except for the four stars with boxes.  These stars
were rejected from the fit as they are $> 5\sigma$ from the mean.
\label{fig_calfac} }
\end{figure*}

\begin{figure}
\epsscale{1.2}
\plotone{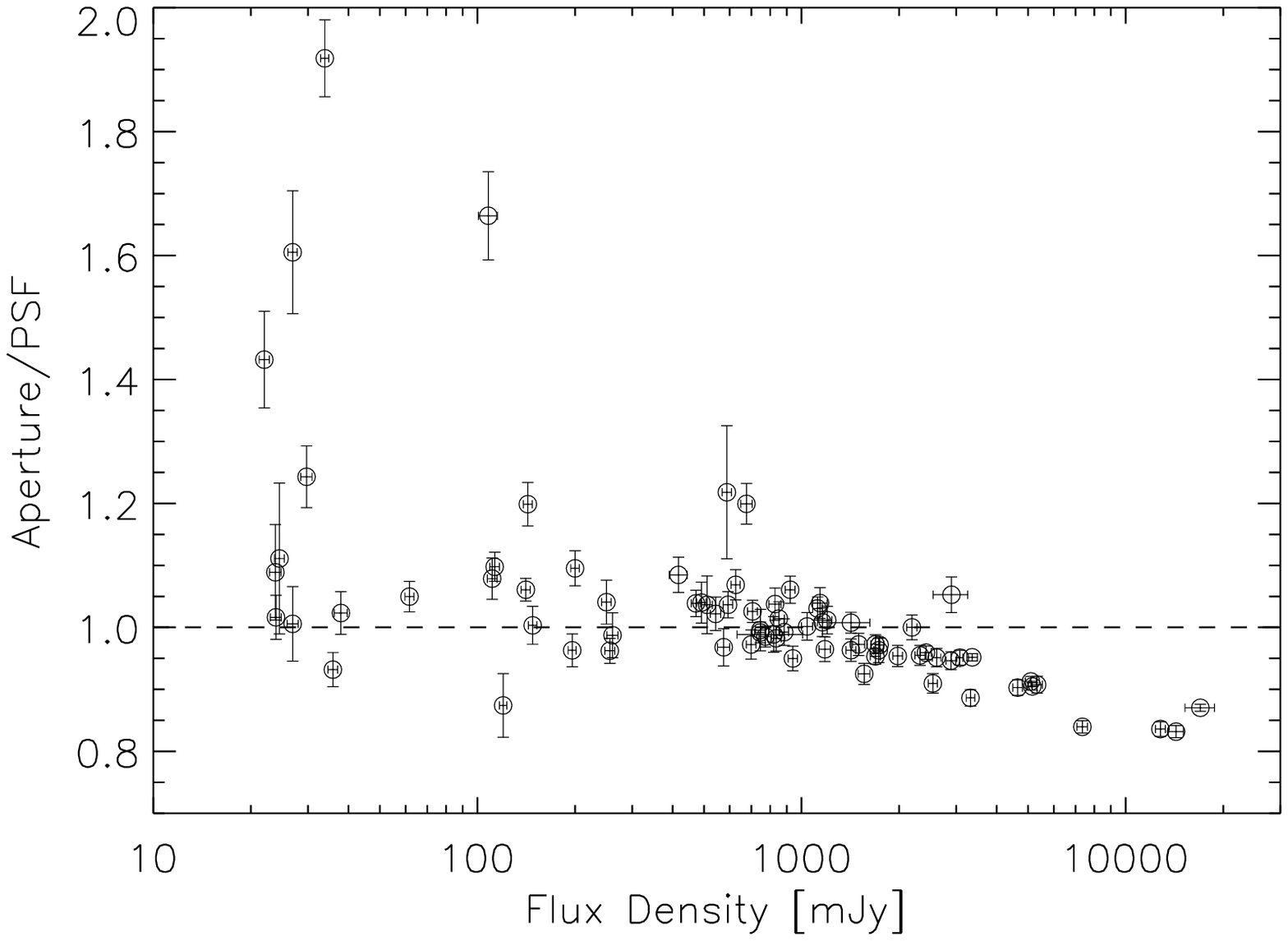}
\caption{The ratio of the aperture to PSF fitting measurements are
given versus predicted flux density.  Each star is represented by a
single point with multiple measurements of a star being averaged
before the ratio computed.
\label{fig_ap_psf_comp} }
\end{figure}

A possible source of scatter in the calibration factor measurements is
flux density nonlinearities that are a characteristic of Ge:Ga
detectors.  In contrast to electronic nonlinearities that depend
solely on total flux density, flux density nonlinearities are due to
the change in rate of flux density falling on the detectors (e.g.,
when a source chops onto a detector).  Given that the
nonlinearity is due to a change in flux density on the detector, the
flux density nonlinearity could depend on background as well as source
flux density.  The electronic nonlinearities
are small ($\sim 2$\%) for the MIPS 70~\micron\ pixels and are
corrected when the data are reduced \citep{Gordon05DAT}.
Fig.~\ref{fig_calfac} shows the calibration factors versus predicted
flux densities and backgrounds.  The calibration factors have been
calculated from the average of all the measurements of each star and
the uncertainty on the average includes the flux density prediction
uncertainty.  The calculated calibration factors for each star are listed in
Tables~\ref{tab_ave_wf} \& \ref{tab_ave_nf}.

It is clear that flux density nonlinearities are present in the
aperture photometry data but are much smaller or non-existent in the
PSF fitting photometry.  A direct comparison of the aperture and PSF
fitting measurements is shown in Fig.~\ref{fig_ap_psf_comp}.  Starting
at around $\sim$1~Jy, the aperture measurements begin to underestimate
the PSF measurements while below $\sim$200~mJy, the aperture to PSF
fitting ratio shows a large scatter.  For bright sources, the flux
density nonlinearities will show up at the peak of the PSF and the PSF
fitting compensates as relatively more weight is given to the wings of
the PSF than in aperture photometry.  For faint sources, the PSF
fitting shows less scatter since it more accurately accounts for nearby
faint sources and background structure than aperture photometry.  For
these reasons, we will use the PSF fitting measurements for the rest
of this work.

The PSF fitting measurements show that the
70~\micron\ calibration measurements are linear 
between 22~mJy and 17~Jy.  A linear fit to these data (accounting for
the uncertainties in both x and y values) gives
\begin{equation}
{\rm Factor} = (693 \pm 18) + (2.9 \pm 5.8)\log(F(\nu))
\end{equation}
where $F(\nu)$ is the flux density in mJy.  There are four stars
(HD 31398, 120933, 213310, \& 216131) that are $> 5\sigma$ from the
mean calibration factor and are identified with square boxes in
Fig.~\ref{fig_calfac}.  The large deviations of these four stars are
most likely due to weak infrared excesses as the derived calibration
factor is low compared to the ensemble.  These four stars have been
excluded from the fit and from the rest of the paper.  The slope value
is consistent with no nonlinearities; the upper limit on the
nonlinearities is computed to be 1.5\% from the ratio of the fit
values at 10~mJy and 20~Jy .  The fit to the aperture photometry
shows a 42\% nonlinearity between the same limits.

A similar result is found for the calibration factor versus
background.  The linear fit to the PSF fitting results gives a slope
of $91 \pm 20$, which is consistent with no slope at the $5\sigma$
level.  There is a statistically significant slope for the linear fit
for the aperture photometry results, but this may just be an artifact of
the nonlinearity seen versus flux density.  A comparison of the two
left hand plots in Fig.~\ref{fig_calfac} shows that the correlation is
better versus flux density than versus background.

\subsection{Repeatability}

\begin{figure}
\epsscale{1.2}
\plotone{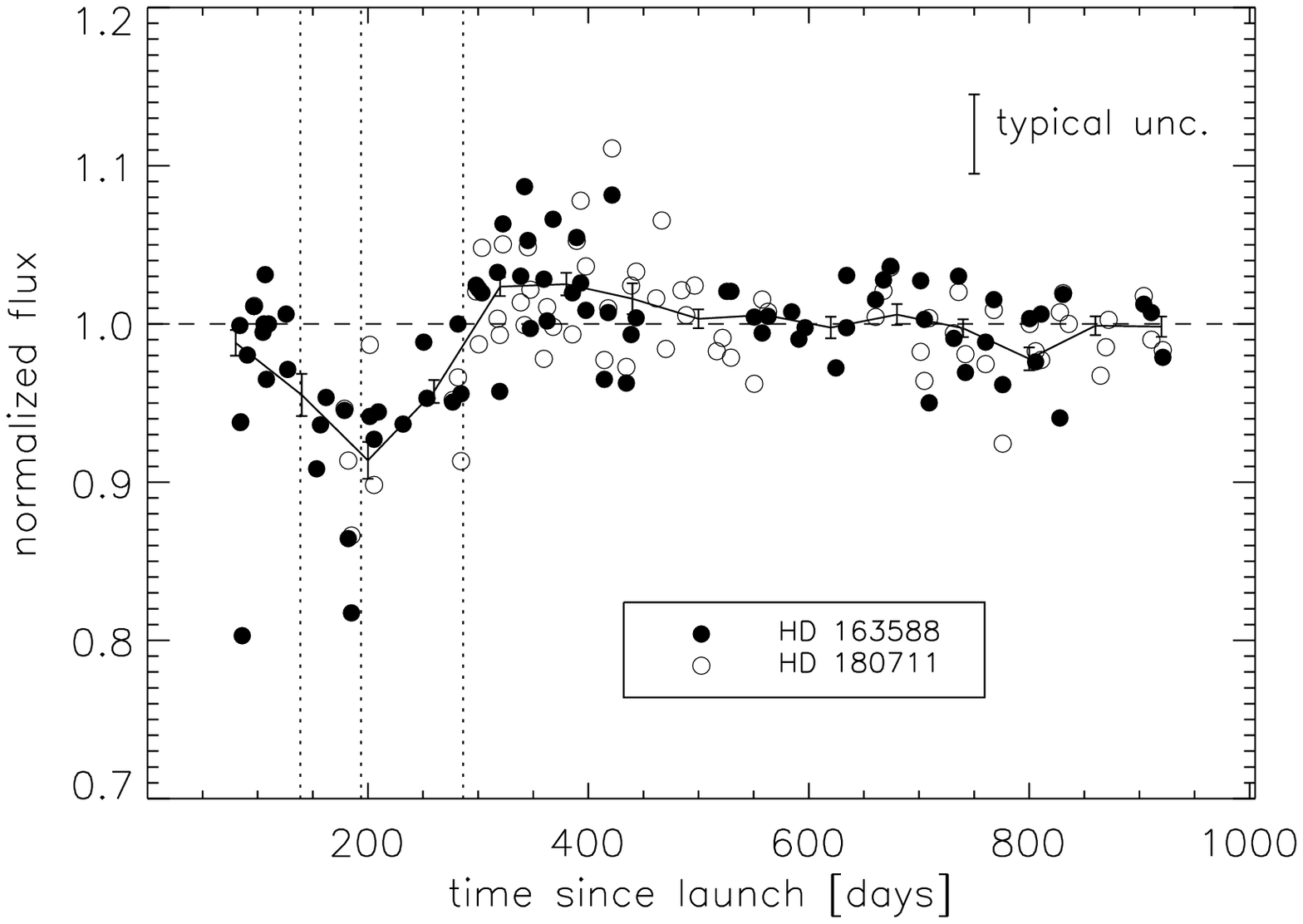}
\caption{The normalized flux densities for the two repeatability
stars, HD~163588 and HD~180711, are plotted versus time.  The solid
line gives the sigma clipped average in 15 equally spaced bins.  The
dashed line is drawn at one.  The typical
uncertainty of a single measurement is shown in the upper right.  The
first vertical line indicates the transition between dither patterns
(between 2nd and 3rd MIPS campaigns).  The second vertical line
indicates the transition between array bias voltage values (between 4th
and 5th MIPS campaigns).  The third vertical line indicates the update
to the instrument software (between the 8th and 9th MIPS campaigns).
\label{fig_calfac_repeat} }
\end{figure}

The photometric repeatability and trends in time can be measured from
the two stars that have been observed throughout MIPS operations.
The photometric repeatability is how well the raw flux of a star can
be measured and is determined from multiple observations of the same
star.  Each measured flux density from PSF fitting photometry was
normalized to the average flux density of each star and plotted in
Fig.~\ref{fig_calfac_repeat}.  The sigma clipped average in 15 bins
between 50 and 950 days since launch is also plotted.  The dotted
vertical lines identify changes in the default dither pattern, bias
voltage, and instrument software.  This plot clearly shows that the
70~\micron\ array displays measurable long term variations.  The
measurements show that the response dropped from the starting value by
$\sim$7\% around day 200 then recovered to slightly above the starting
value around day 300.  After this, the response seems to have
stabilized with evidence for a weak trend downward.  While it is
tempting to identify the initial variations with the changes in how
the data were taken, no clear instrumental parameter has been
identified that would cause these variations.  Initial testing with
other methods of measuring the brightness of these two stars has shown
similar, but not identical variations.  More work is clearly needed to
understand the origin of the initial variations.

Even given the initial variations, the repeatability of the two
stars is quite good.  The repeatabilities for all the measurements are
4.5\% for HD~163588 and 3.7\% for HD~180711.  For reference, the
repeatabilities for aperture photometry using the same data are 4.9\%
for HD~163588 and 3.9\% for HD~180711.  Given the changes in the
operating parameters of the 70~\micron\ array early in the mission, we
also computed the repeatabilities using only the data taken after the
last change.  The repeatabilities for all the measurements after the
8th MIPS campaign are 2.9\% for HD~163588 and 2.7\% for HD~180711.
For reference, the repeatabilities for aperture photometry using the
same data are 4.3\% for HD~163588 and 3.4\% for HD~180711.  The
combined measurements imply a conservative repeatability of the MIPS
70~\micron\ array of $\sim$4.5\%.

\subsection{Time Since Anneal}

\begin{figure}
\epsscale{1.2}
\plotone{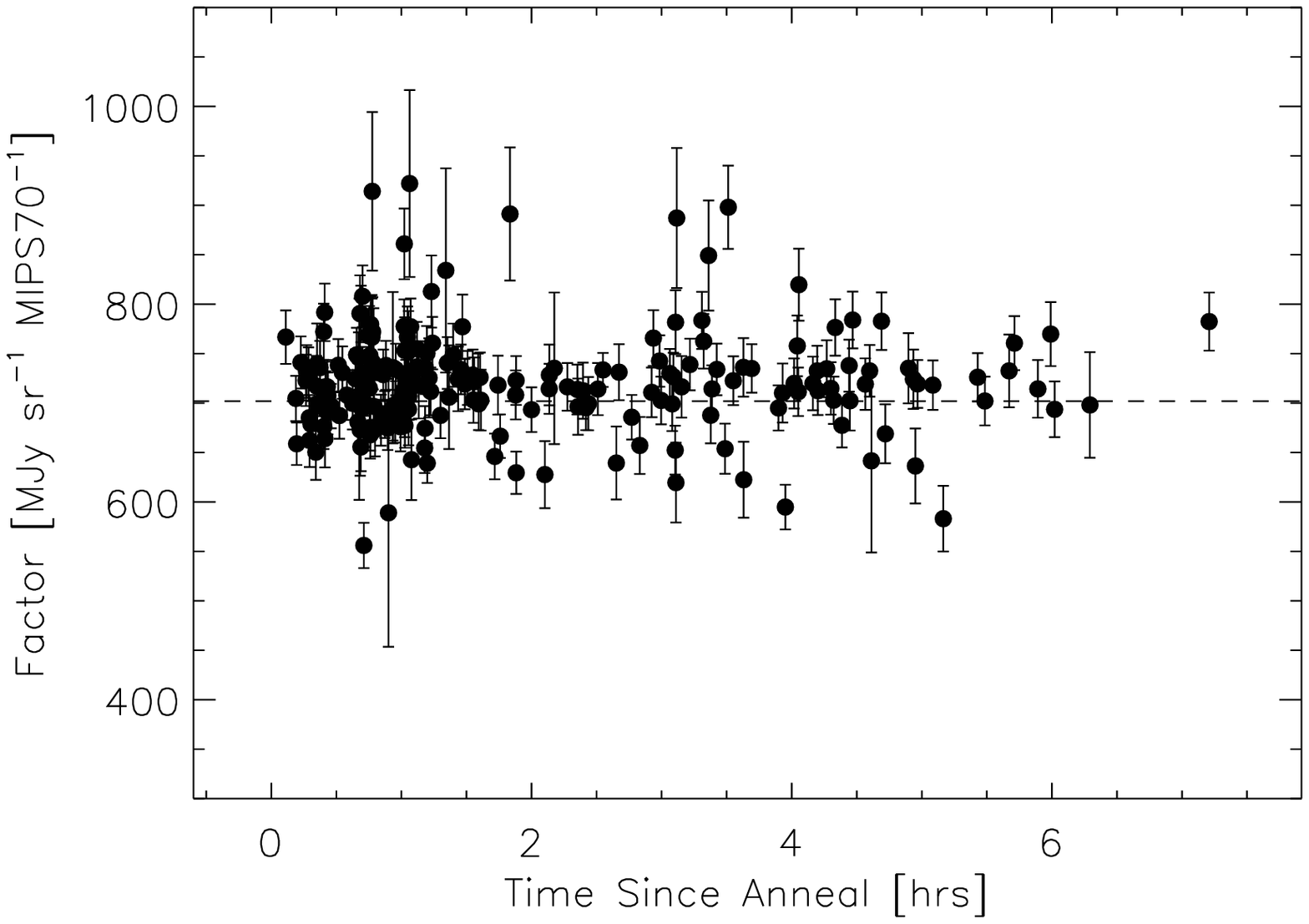}
\caption{The calibration factor is plotted versus time since anneal
for all the measurements.  The dashed line is drawn at the final
calibration factor.
\label{fig_calfac_atime} }
\end{figure}

The MIPS 70~\micron\ array is annealed by raising the temperature by a
few degrees to remove cosmic ray damage.  This was done every six
hours till MIPS campaign 20 and every three to four hours since
then.  Residual instrumental signatures are seen to grow with time
since anneal and this change was made to minimize them.  In addition
to removing cosmic ray damage, the responsivity of the array is reset
\citep{Rieke04MIPS}.  The calibration factor should not have a
dependence on time since anneal as all 70~\micron\ measurements are
referenced to the internal simulator measurements taken every two
minutes or less.  To check this, Figure~\ref{fig_calfac_atime} shows
the calibration factor versus anneal time.  No trend is seen.

\subsection{Other Checks}

The accuracy of the flux density predictions can be checked by
comparing the calibration factor derived for stars of different
spectral types.  The stars in the sample can be divided into three
categories; hot stars (B \& A dwarfs), solar analogs (early G dwarfs),
and cool stars (K \& M giants).  The weighted average calibration
factors (after sigma clipping) for these three categories are $717 \pm
8$, $717 \pm 3$, and $701 \pm 6$ MJy sr$^{-1}$ MIPS70$^{-1}$ where the
number of measurements contributing to each class are 11, 6, and 58,
respectively.  None of averages are significantly different from each
other, especially
when the small number of measurements contributing to the hot stars
and solar analogs are taken into account.

The dependence on exposure time was checked by computing the weighted
average calibration factor separately for observations taken with 3.15
and 10.49~s.  The resulting calibration factors are $701 \pm 6$ and
$723 \pm 7$ MJy sr$^{-1}$ MIPS70$^{-1}$ with 62 and 12
measurements contributing to the averages, respectively.  As the
difference is only $2.4\sigma$, no
significant systematic 
change with exposure time is seen.

\subsection{Noise Characteristics}
\label{sec_noise}

\begin{figure}
\epsscale{1.2}
\plotone{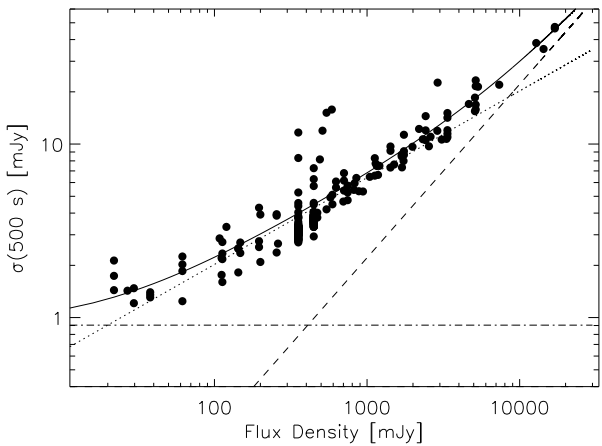}
\caption{The noise is plotted versus predicted flux density for the
PSF fitting measurements.
Each noise measurement has been converted to the equivalent noise in
500~s assuming the noise scales as time$^{0.5}$.  The solid line gives
the fit to the data that is discussed in
$\S$\ref{sec_noise}.  The dot-dashed line gives the constant term from
the fit, the dotted line the term proportional to F$(\nu)^{0.5}$, and
the dashed line the term proportional to F$(\nu)$. \label{fig_ynoise} }  
\end{figure}

The behavior of the noise is plotted versus predicted flux density in
Fig.~\ref{fig_ynoise} for the PSF fitting measurements.  To
compare measurements taken at different exposure times, each noise
measurement has been transformed to the equivalent noise in 500~s
assuming a $t^{0.5}$ dependence.  The noise behavior can be
characterized by 
\begin{equation}
\sigma(\mathrm{500 s})^2 = 0.90^2 + [0.20 \mathrm{F(\nu)}^{0.5}]^2 + [0.0022 \mathrm{F(\nu)}]^2
\end{equation}
where $F(\nu)$ is the predicted 70~\micron\ flux density in mJy.  
The
first term accounts for the confusion noise, the second term the
photon noise, and the third term the noise due to the division by the
interpolated stimulator flash.

The sensitivity of the MIPS 70~\micron\ band in 500~s can be
determined from this fit by computing where the flux is 5$\times$ the
uncertainty from Eq. (2). The  5$\sigma$, 500~s sensitivity computed in
this fashion is $\sim$5~mJy.
This measurement is based on an extrapolation of over a factor of ten
from the lowest measured point and so is fairly uncertain.  The
sensitivity can be better measured from deep cosmological surveys.
The Extragalactic First Look Survey gives a 5$\sigma$, 500~s
sensitivity of $\sim$6~mJy after correcting for the updated
calibration factor \citep{Frayer06}.  Somewhat worse
sensitivities up to $\sim$8~mJy are seen for other deep MIPS
70~\micron\ cosmological fields (D.\ Frayer, C.\ Papovich, private
communication).  This sensitivity can include a contribution from
confusion, but this is likely to be small since the 5$\sigma$
confusion noise for MIPS 70~\micron\ coarse-scale observations is
estimated at $\sim$1.5--3~mJy
\citep{Dole04b,Frayer06b}.  Combining the results from the calibration
stars and deep cosmological field gives a conservative
5$\sigma$, 500~s sensitivity of 6--8~mJy.

\subsection{Coarse-Scale Calibration}
\label{sec_calfac_wf}

The final coarse-scale calibration is based on the PSF fitting results
from all the observations (top, right plot in Fig.~\ref{fig_calfac}).
The final calibration factor of $702 \pm
35$~MJy~sr$^{-1}$~MIPS70$^{-1}$ was determined by performing a weighted average
of the calibration factors measured for each star.  There are 66 calibration stars that
contribute to this calibration factor.  The calibration accuracy is
dominated by the 4.5\% repeatability uncertainty, but also includes
contributions from the uncertainty of the mean (very small) and the
2\% systematic uncertainty in the 24/70~\micron\ flux density ratios.
The conversion directly to flux densities from MIPS70 units
implied by the measured calibration factor is $1.60 \pm
0.08$~Jy~MIPS70$^{-1}$ given the instrumental flux densities were
measured on mosaics with $9.85\arcsec \times 9.85\arcsec$ pixels.
The calibration factor determined from the 57 stars taken after the 9th
MIPS campaign is $696 \pm 34$~MJy~sr$^{-1}$~MIPS70$^{-1}$.  This shows
that the calibration factor is not significantly changed by the
variations in the repeatability stars seen prior to the 9th MIPS
campaign.  The calibration factor determined from the aperture
measurements is $704 \pm 35$~MJy~sr$^{-1}$~MIPS70$^{-1}$ where only
measurements from 0.3 to 1~Jy
are used (see Figs.~\ref{fig_calfac} \& \ref{fig_ap_psf_comp}).
This shows that the calibration derived from aperture and PSF fitting
measurements are equivalent in the restricted range where the aperture
photometry produces accurate results.

The preliminary calibration factor determined soon after launch was
$634 \pm 127$~MJy~sr$^{-1}$~MIPS70$^{-1}$.  The new value is 11\%
larger than this preliminary value, well within the 20\% uncertainty
in the preliminary value.  This 11\% change is not unexpected given
that the preliminary value was based on a handful of measurements of a
single star and the data reduction at the time did not include the
extra steps to correct residual instrumental signatures.  Existing
data can be corrected to the new calibration factor by multiplying by
the ratio of the new to old calibration factors.  Note the calibration
factor applied to data reduced using the SSC pipeline is given by the
FITS header keyword FLUXCONV.

The coarse-scale calibration factor is determined from photometry mode
observations, but should apply directly to scan map mode observations
as both modes share the same optical train and only differ in the
dithering strategy.  The relative response between the scan map and
coarse-scale photometry mode has been checked by observing the same
source in both modes.  It was found that the same flux density was
measured within the uncertainties.

The consistency of the 70~\micron\ calibration with the 24~\micron\
calibration was checked using stars that were observed at both 24 and
70~\micron.  There are 36 such measurements for stars with 24~\micron\
flux
densities below 4~Jy (24~\micron\ saturation limit).  The resulting
sigma clipped average of the ratio of the observed to model 24/70
color ratio is 
$1.002 \pm 0.013$ showing that the 24 and 70~\micron\ calibrations are
consistent.

\subsection{Extended Versus Point Source Calibration Check}

\begin{figure}
\epsscale{1.2}
\plotone{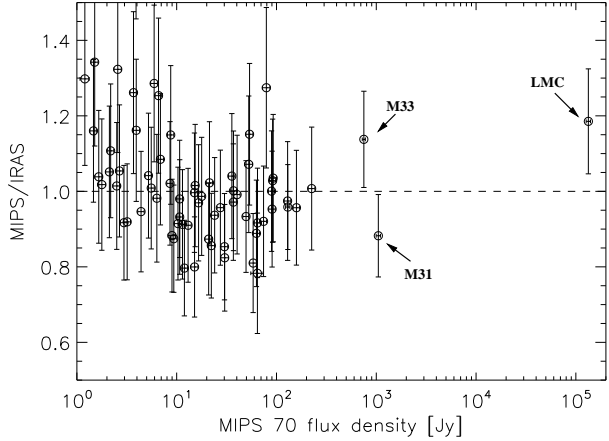}
\caption{The ratio of MIPS 70~\micron\ measured flux density to the
70~\micron\ flux density predicted from IRAS 60 and 100~\micron\
measurements is shown as a check of the extended source calibration.
The sample shown is comprised of the 75 SINGS galaxies supplemented by
the LMC, M31, M33, and M101.  The uncertainties are computed from the
uncertainties in the IRAS and MIPS measurements.  The dashed line is
drawn at a value of one.
\label{fig_extcheck} }
\end{figure}

The calibration factor determined above was calculated from
observations and predictions of point sources.  Given the complex
response of Ge:Ga detectors to sources with different spatial extents,
it is important to verify that the point source calibration applies to
extended sources.  This can be checked by comparing the total fluxes
of resolved galaxies measured at 70~\micron\ to those predicted by
Infrared Astronomical Satellite \citep[IRAS,][]{Beichman88}
measurements at 60 and 100~\micron.  Galaxies provide good objects for
such a check as they are discrete extended sources that were well
measured by IRAS and have a significant component of their flux
that is resolved by MIPS at 70~\micron.

The comparison is done using the 75 Spitzer Infrared Nearby Galaxies
Survey \citep[SINGS,][]{Kennicutt03SINGS} galaxies from global
measurements given by \citet{Dale05} and updated by \citet{Dale07}.
This sample is supplemented at the highest flux densities 
with global measurements of M31 \citep{Gordon06}, M33 \citep{Hinz04},
M101 (Gordon et al. in prep.), and the Large Magellanic Cloud
\citep[LMC,][]{Meixner06}.  The predictions of the 70~\micron\ flux
densities from the IRAS 60 and 100~\micron\ measurements were
done by first color correcting the IRAS measurements using the
measured 60/100 flux density ratio to pick the appropriate power law
color correction \citep{Beichman88} and then interpolating to the
effective wavelength of the MIPS 70~\micron\ band of 71.42~\micron.
The average color corrections were 1.0 for both IRAS 60 and 100 bands.
The MIPS 70~\micron\ measurements were corrected to the updated
calibration factor and color corrected using the correction for the
same power law determined for the IRAS measurements
\citep{Stansberry07MIPS160}.  The average MIPS 70~\micron\ color
correction was 0.93.  Figure~\ref{fig_extcheck} gives the ratio of the
MIPS to predicted IRAS 70~\micron\ flux densities for all the galaxies
with flux densities above 1~Jy.  The weighted average of this ratio is
0.99 which is well within the uncertainties on the absolute
calibration of MIPS 70~\micron\ (5\%) and IRAS 60~\micron\
\citep[5\%,][]{Beichman88}.  This shows that the the MIPS
70~\micron\ point source calibration applies for extended sources.
This comparison also serves as another check that the photometry and
scan observing modes share the same calibration even through the
calibration factor is derived from photometry mode observations while
the galaxies were observed with the scan map mode.

\subsection{Fine-Scale Calibration}

\begin{figure}
\epsscale{1.2}
\plotone{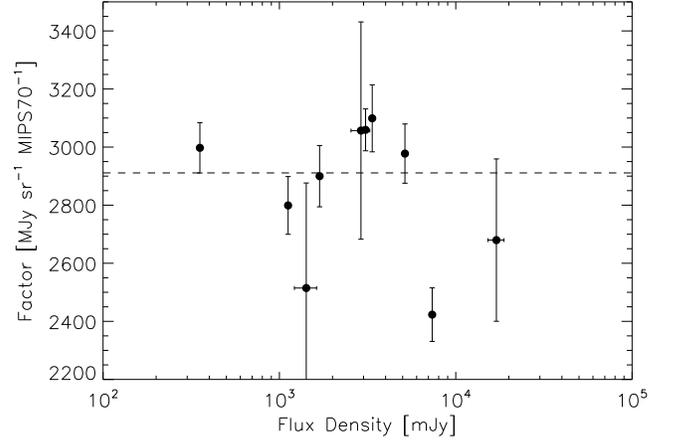}
\caption{The fine-scale calibration factor is plotted versus predicted
flux density.  The multiple measurements of each star have been
averaged and the plotted uncertainty includes the flux density
prediction uncertainty. The final calibration factor is shown as a
dashed line.
\label{fig_calfac_nf} }
\end{figure}

The fine-scale calibration should be similar to the coarse-scale
calibration modulo the different pixel scales (5.24 instead of
9.85\arcsec\ pixel$^{-1}$) and optical trains.
Fig.~\ref{fig_calfac_nf} shows the 
calibration factor for PSF fitting photometry.  The fine-scale calibration factor
determined from these data is $2894 \pm
294$~MJy~sr$^{-1}$~MIPS70F$^{-1}$.  The conversion directly
to flux densities from MIPS70F units 
implied by this calibration factor is $1.87 \pm
0.188$~Jy~MIPS70$^{-1}$ given the instrumental flux densities were
measured on mosaics with $5.24\arcsec \times 5.24\arcsec$ pixels.
The fine-scale calibration factor
is 4.12$\times$ the coarse-scale factor.  This is larger than the
ratio of pixel areas (3.53) implying that the different optical train
has a significant effect on the calibration.  Only 10 calibration stars
are used for the fine-scale calibration and, as a result, the
fine-scale calibration must be taken as preliminary.  The 
uncertainty on the fine-scale calibration is formally only 5\%, but we
have doubled this to account for the small sample used.  In
general, the fine-scale mode should be used for probing structure and
the coarse-scale used when accurate photometry is needed.

\subsection{Comparison to Previous Missions}

The repeatability, absolute calibration accuracy, and sensitivity of
the coarse-scale MIPS 70~\micron\ band can be compared to previous
missions.  Imaging of point sources in bands similar to the MIPS
70~\micron\ band has been provided in the past by the IRAS and the
Infrared Space Observatory \citep[ISO,][]{Kessler96}.

IRAS 60~\micron\ photometry of points sources has a repeatability of
11\% and an absolute calibration uncertainty of 5\%
\citep{Beichman88}.  The IRAS Faint Source Catalog \citep{Moshir92} is
94\% complete at 0.2~Jy.  The almost full sky coverage of IRAS allows
for a comparison of the IRAS 60~\micron\ and MIPS 70~\micron\ fluxes.
There are 30 stars with both MIPS 70~\micron\ and IRAS 60~\micron\
Faint Source Catalog 
measurements \citep{Moshir90} and IRAS 60~\micron\ fluxes above
0.6~Jy.  The average 
ratio of the IRAS 60~\micron\ to MIPS 70~\micron\ flux densities is
$1.35 \pm 0.01$ and the expected ratio from a Rayleigh-Jeans model is
1.42.  Thus, the IRAS 60~\micron\ flux densities need to be multiplied
by 1.05 in order to put them on the \citet{Rieke07absir} system.

The \citep[ISOPHOT,][]{Laureijs03} instrument included a 70~\micron\
band with the P3 detector and a 60~\micron\ band with the C100
detector.  The P3 reproducibility was 10-20\% and absolute accuracy
was 7\%.  The C100 repeatability was 3-20\% and absolute accuracy
was 15-25\% \citep{Klaas03}.  The sensitivity of ISOPHOT in these
bands is approximately 7.5-20~mJy $1\sigma$ in 256~s
\citep{Laureijs03} which corresponded to a 5$\sigma$, 500~s
sensitivity of 30-70~mJy.  This sensitivity is better than that found
from the 95~\micron\ deep field observations of \citet{Rodighiero03}
where the 3$\sigma$, 5184~s sensitivity is 16~mJy which corresponds to
a 5$\sigma$, 500~s sensitivity of 86~mJy.

The MIPS 70~\micron\ observations compare favorably, with better
repeatability (4.5\%), as good or better absolute calibration
uncertainty (5\%), and higher sensitivity (6-8~mJy 5$\sigma$,
500~s) than previous missions imaging in similar bands.  This is to be
expected as the MIPS 70~\micron\ detector operation has benefited
from lessons learned from past missions.

\section{Summary}

The calibration of the MIPS 70~\micron\ coarse- and fine-scale imaging
modes was determined from many observations taken over the first 2.5
years of the Spitzer mission.

\begin{enumerate}

\item The characterization and calibration of the MIPS 70~\micron\
coarse-scale mode was determined from measurements of 78 stars with
spectral types from B to M, with flux densities from 22~mJy to 17~Jy,
and on backgrounds from 4 to 26 MJy sr$^{-1}$.  The coarse-scale
calibration factor is $702 \pm 
35$~MJy~sr$^{-1}$~MIPS70$^{-1}$ and was determined from measurements
of 66 stars.  A handful of stars were rejected due to possible
infrared excesses, contamination from nearby extended emission, and
nearby very bright sources.

\item Accurate photometry of point sources in coarse-scale mode
requires two simple 
processing steps beyond the standard data reduction to remove
long-term detector transients.

\item PSF fitting photometry is seen to produce better measurements than
aperture photometry due to better handling of nearby stars, background
structure, and less weighting of the PSF core where flux
non-linearities have a larger effect.

\item Using PSF fitting photometry, no significant trends in
calibration factor versus predicted flux density, predicted background,
exposure time, spectral type, and time since anneal were found.

\item The photometric
repeatability is 4.5\% measured from two stars observed during every
campaign and includes variations on all time scales probed.

\item The 5$\sigma$, 500~s sensitivity of coarse-scale observations is
6--8~mJy and was determined from the calibration stars and deep
cosmological surveys.

\item The applicability of the coarse-scale calibration factor,
derived from point source observations, to extended sources was
confirmed using a sample of galaxies observed with MIPS and IRAS.

\item The preliminary fine-scale calibration factor is $2894 \pm
294$~MJy~sr$^{-1}$~MIPS70F$^{-1}$ and was determined from measurements
of 10 stars with flux densities from 350~mJy to 17~Jy.  

\end{enumerate}

\acknowledgements
We thank the anonymous referee for comments which improved the paper.
This work is based on observations made with the {\em Spitzer Space
Telescope}, which is operated by the Jet Propulsion Laboratory,
California Institute of Technology under NASA contract 1407. Support
for this work was provided by NASA through Contract Number \#1255094
issued by JPL/Caltech.

\LongTables
\clearpage

\begin{deluxetable}{lccccc}
\tablewidth{0pt}
\tablecaption{Aperture Corrections \label{tab_apcor}}
\tablehead{ \colhead{Description} & \colhead{Radius} & 
   \colhead{Background} & \colhead{T=10,000K} &
   \colhead{T=60K} & \colhead{T=10K} \\
  & \colhead{[$\arcsec$]} & \colhead{[$\arcsec$]} &
    \colhead{PSF} & \colhead{PSF} & \colhead{PSF}}
\startdata
\multicolumn{6}{c}{Coarse-Scale} \\ \hline
$>$2nd Airy Ring & 100 & 120-140 & 1.10 & 1.10 & 1.13 \\
$>$1st Airy Ring &  35 &   39-65 & 1.22 & 1.24 & 1.48 \\
   2$\times$HWHM &  16 &   18-39 & 2.04 & 2.07 & 2.30 \\ \hline
\multicolumn{6}{c}{Fine-Scale} \\ \hline
$>$2nd Airy Ring & 100 & 120-140 & 1.10 & 1.10 & 1.13 \\
$>$1st Airy Ring &  35 &   39-65 & 1.21 & 1.22 & 1.47 \\
   2$\times$HWHM &  16 &   18-39 & 1.93 & 1.94 & 2.16 \\
\enddata
\end{deluxetable}

\clearpage

\begin{deluxetable}{lrrrrrrrrr}
\tablewidth{0pt}
\tablecaption{Coarse-Scale Measurements \label{tab_meas_wf}}
\tablehead{\colhead{Name} & \colhead{Campaign\tablenotemark{a}} &
    \colhead{AOR \#} & \colhead{Time\tablenotemark{b}} & 
    \colhead{ExpTime} & \colhead{TotTime\tablenotemark{c}} &
    \colhead{Ap. Flux} & \colhead{S/N} & \colhead{PSF Flux} & \colhead{S/N} \\
 & & & \colhead{[s]} & \colhead{[s]} & \colhead{[s]} & 
    \colhead{[MIPS70]} & & \colhead{[MIPS70]} & }
\startdata
HD002151 & 8MC & 9806592 & 770515712 & 10.49 & 259.5 & 1.53E$-$01 & 20.4 & 1.62E$-$01 & 47.6 \\
 & 10MC & 10091520 & 773862336 & 3.15 & 77.7 & 1.52E$-$01 & 20.6 & 1.58E$-$01 & 42.5 \\
 & 26MC & 16276992 & 815940416 & 3.15 & 78.0 & 1.52E$-$01 & 22.9 & 1.52E$-$01 & 25.6 \\
HD002261 & X1 & 7977216 & 753515520 & 3.15 & 78.3 & 7.15E$-$01 & 93.5 & 6.88E$-$01 & 58.3 \\
HD003712 & 11MC & 11784448 & 775613056 & 3.15 & 78.1 & 7.11E$-$01 & 135.3 & 7.37E$-$01 & 70.2 \\
HD004128 & 16MC & 12873216 & 786539520 & 3.15 & 78.2 & 6.98E$-$01 & 143.9 & 6.90E$-$01 & 63.6 \\
HD006860 & 12MC & 11893504 & 777752832 & 3.15 & 78.5 & 3.04E+00 & 209.6 & 3.35E+00 & 98.8 \\
HD009053 & 11MC & 11784960 & 775614464 & 3.15 & 78.2 & 8.85E$-$01 & 201.3 & 9.10E$-$01 & 78.0 \\
HD009927 & 11MC & 11784704 & 776018752 & 3.15 & 78.3 & 2.93E$-$01 & 85.2 & 2.77E$-$01 & 49.6 \\
 & 28MC & 16619776 & 821558080 & 3.15 & 77.8 & 2.98E$-$01 & 79.8 & 2.94E$-$01 & 46.3 \\
HD012533 & 12MC & 11893760 & 777753856 & 3.15 & 78.5 & 1.24E+00 & 189.0 & 1.30E+00 & 81.5 \\
HD012929 & 18MC & 13113344 & 791238208 & 3.15 & 53.2 & 1.04E+00 & 144.7 & 1.08E+00 & 70.7 \\
HD015008 & 13MC & 12064768 & 780277312 & 10.49 & 510.9 & 1.98E$-$02 & 9.2 & 1.36E$-$02 & 15.4 \\
 & 13MC & 12154368 & 780623616 & 10.49 & 511.0 & 1.82E$-$02 & 9.5 & 1.04E$-$02 & 10.4 \\
 & 13MC & 12154624 & 780621184 & 10.49 & 510.3 & 1.50E$-$02 & 9.4 & 1.31E$-$02 & 12.8 \\
HD018884 & 12MC & 11894016 & 777747392 & 3.15 & 78.5 & 2.60E+00 & 268.4 & 2.88E+00 & 108.2 \\
HD020902 & 29MC & 16868864 & 824522048 & 10.49 & 260.8 & 3.13E$-$01 & 87.3 & 3.02E$-$01 & 30.9 \\
HD024512 & 1MC & 8358144 & 755758720 & 10.49 & 91.2 & 1.29E+00 & 175.2 & 1.36E+00 & 71.4 \\
 & 1MC & 8358400 & 755759040 & 3.15 & 52.6 & 1.47E+00 & 173.5 & 1.54E+00 & 74.6 \\
 & 14MC & 12197376 & 782166144 & 3.15 & 78.3 & 1.46E+00 & 211.6 & 1.52E+00 & 79.9 \\
HD025025 & 13MC & 12063744 & 779604736 & 3.15 & 78.2 & 1.48E+00 & 223.5 & 1.55E+00 & 86.1 \\
HD029139 & 19MC & 13315072 & 793954560 & 3.15 & 53.1 & 5.96E+00 & 174.9 & 7.13E+00 & 109.5 \\
HD031398 & 13MC & 12064000 & 780105728 & 3.15 & 78.3 & 1.11E+00 & 176.1 & 1.20E+00 & 79.3 \\
HD032887 & 13MC & 12064256 & 779604352 & 3.15 & 78.1 & 7.31E$-$01 & 172.4 & 7.25E$-$01 & 61.3 \\
HD034029 & 4MC & 9059840 & 761915840 & 3.15 & 78.5 & 2.74E+00 & 331.2 & 2.91E+00 & 130.5 \\
 & 13MC & 12064512 & 780235008 & 3.15 & 78.5 & 2.69E+00 & 240.4 & 3.08E+00 & 109.0 \\
HD035666 & 9MC & 9941248 & 772043584 & 10.49 & 510.9 & 2.07E$-$02 & 14.8 & 1.76E$-$02 & 16.8 \\
 & 20MC & 13478656 & 797749696 & 3.15 & 304.7 & 1.47E$-$02 & 7.0 & 1.76E$-$02 & 16.5 \\
HD036167 & 19MC & 13308416 & 794234816 & 3.15 & 78.0 & 2.42E$-$01 & 27.5 & 2.34E$-$01 & 37.4 \\
 & 29MC & 16869120 & 825753280 & 10.49 & 259.9 & 2.72E$-$01 & 30.4 & 2.39E$-$01 & 35.8 \\
HD039425 & R & 7600896 & 753649280 & 3.15 & 78.1 & 3.52E$-$01 & 58.0 & 2.89E$-$01 & 14.7 \\
HD039608 & 12MC & 11892992 & 777732992 & 10.49 & 510.8 & 3.09E$-$02 & 21.3 & 2.23E$-$02 & 24.8 \\
 & 20MC & 13479168 & 797750528 & 10.49 & 510.9 & 2.14E$-$02 & 12.4 & 2.02E$-$02 & 20.3 \\
HD042701 & 9MC & 9941504 & 772044608 & 10.49 & 510.7 & 6.10E$-$02 & 43.8 & 3.18E$-$02 & 34.8 \\
HD045348 & 6MC & 9459712 & 765979776 & 3.15 & 78.2 & 1.75E+00 & 317.7 & 1.84E+00 & 114.9 \\
HD048915 & 6MC & 9458432 & 765977856 & 3.15 & 78.4 & 1.60E+00 & 246.8 & 1.52E+00 & 50.8 \\
HD050310 & X1 & 7977984 & 753518784 & 3.15 & 78.0 & 4.13E$-$01 & 52.3 & 4.14E$-$01 & 41.0 \\
 & X1 & 7979520 & 753518528 & 3.15 & 78.3 & 4.49E$-$01 & 61.7 & 4.28E$-$01 & 45.4 \\
 & 21MC & 13641472 & 800741248 & 3.15 & 78.4 & 4.45E$-$01 & 105.3 & 4.35E$-$01 & 51.9 \\
HD051799 & R & 7601152 & 753649664 & 3.15 & 71.5 & 4.09E$-$01 & 84.2 & 3.93E$-$01 & 44.0 \\
 & 21MC & 13641728 & 800740928 & 3.15 & 78.2 & 3.90E$-$01 & 113.9 & 3.79E$-$01 & 52.2 \\
HD053501 & R & 7601408 & 753650112 & 3.15 & 77.8 & 9.13E$-$02 & 13.4 & 8.35E$-$02 & 22.6 \\
 & X1 & 7977728 & 753518144 & 3.15 & 77.9 & 1.14E$-$01 & 15.6 & 1.00E$-$01 & 22.7 \\
 & 21MC & 13641984 & 800744576 & 3.15 & 78.2 & 9.84E$-$02 & 25.3 & 7.73E$-$02 & 31.1 \\
HD056855 & 19MC & 13315328 & 793945152 & 3.15 & 53.1 & 1.61E+00 & 187.3 & 1.77E+00 & 85.4 \\
HD059717 & 15MC & 12397824 & 784145600 & 3.15 & 78.3 & 8.89E$-$01 & 159.8 & 9.23E$-$01 & 76.9 \\
HD060522 & 20MC & 13440768 & 797687296 & 3.15 & 52.8 & 4.54E$-$01 & 92.8 & 4.56E$-$01 & 42.1 \\
HD062509 & 15MC & 12398080 & 784147584 & 3.15 & 78.6 & 1.56E+00 & 171.9 & 1.64E+00 & 93.7 \\
HD071129 & 3MC & 8813312 & 759211520 & 3.15 & 78.4 & 3.12E+00 & 264.5 & 3.08E+00 & 87.6 \\
 & 16MC & 12873472 & 786542720 & 3.15 & 78.6 & 2.83E+00 & 201.5 & 3.29E+00 & 129.2 \\
 & 28MC & 16620032 & 821426816 & 3.15 & 78.5 & 2.75E+00 & 197.1 & 3.13E+00 & 121.0 \\
 & 29MC & 16868608 & 824534592 & 3.15 & 78.5 & 2.81E+00 & 204.9 & 2.99E+00 & 94.7 \\
HD080007 & 6MC & 9459200 & 765981056 & 10.49 & 259.7 & 1.34E$-$01 & 38.4 & 1.23E$-$01 & 36.7 \\
 & 19MC & 13308672 & 794194496 & 3.15 & 78.1 & 1.42E$-$01 & 36.1 & 1.30E$-$01 & 37.8 \\
HD080493 & 21MC & 13634304 & 800483200 & 3.15 & 78.3 & 1.02E+00 & 166.4 & 1.08E+00 & 78.6 \\
 & 21MC & 13642752 & 800733696 & 3.15 & 78.4 & 1.03E+00 & 195.9 & 1.06E+00 & 77.0 \\
HD081797 & 21MC & 13634048 & 800449280 & 3.15 & 78.4 & 1.76E+00 & 212.0 & 1.86E+00 & 95.9 \\
HD082308 & W & 7966464 & 754628224 & 3.15 & 78.2 & 3.58E$-$01 & 43.2 & 2.68E$-$01 & 14.2 \\
 & 21MC & 13643264 & 800736128 & 3.15 & 78.1 & 3.25E$-$01 & 79.5 & 3.27E$-$01 & 51.0 \\
HD082668 & 7MC & 9661952 & 768555648 & 3.15 & 78.1 & 9.61E$-$01 & 107.6 & 9.73E$-$01 & 61.5 \\
 & 10MC & 10091264 & 773861824 & 3.15 & 78.1 & 9.13E$-$01 & 122.7 & 8.87E$-$01 & 58.1 \\
HD087901 & W & 7965440 & 754625728 & 3.15 & 78.0 & 1.05E$-$01 & 14.1 & 1.11E$-$01 & 28.1 \\
 & 7MC & 9661440 & 767950848 & 10.49 & 261.1 & 1.02E$-$01 & 18.8 & 9.98E$-$02 & 33.0 \\
 & 7MC & 9662720 & 767992960 & 3.15 & 78.1 & 1.07E$-$01 & 17.7 & 1.20E$-$01 & 30.4 \\
HD089388 & 20MC & 13441024 & 797793856 & 3.15 & 52.9 & 6.04E$-$01 & 68.1 & 6.11E$-$01 & 49.0 \\
HD089484 & W & 7967232 & 754630208 & 3.15 & 78.3 & 1.08E+00 & 154.8 & 1.09E+00 & 61.0 \\
 & 21MC & 13634560 & 800463488 & 3.15 & 78.4 & 1.01E+00 & 186.8 & 1.05E+00 & 79.7 \\
 & 21MC & 13643008 & 800736512 & 3.15 & 78.2 & 1.02E+00 & 181.9 & 1.05E+00 & 75.8 \\
HD089758 & W & 7967488 & 754630592 & 3.15 & 77.9 & 1.37E+00 & 130.6 & 1.37E+00 & 70.8 \\
HD092305 & 19MC & 13308928 & 794193536 & 3.15 & 78.0 & 5.01E$-$01 & 120.3 & 5.03E$-$01 & 62.1 \\
 & 21MC & 13590272 & 800828224 & 3.15 & 78.1 & 4.60E$-$01 & 125.1 & 4.66E$-$01 & 56.1 \\
HD093813 & 22MC & 15247872 & 803701696 & 3.15 & 77.9 & 4.57E$-$01 & 53.6 & 4.70E$-$01 & 59.2 \\
HD095689 & 21MC & 13634816 & 800476096 & 3.15 & 78.2 & 1.07E+00 & 162.4 & 1.10E+00 & 91.3 \\
HD096833 & W & 7966720 & 754628672 & 3.15 & 77.9 & 3.81E$-$01 & 50.2 & 3.62E$-$01 & 40.7 \\
 & 28MC & 16619008 & 821403072 & 3.15 & 78.3 & 3.95E$-$01 & 78.3 & 3.66E$-$01 & 44.1 \\
HD100029 & X1 & 7980032 & 753522304 & 3.15 & 78.2 & 7.76E$-$01 & 87.0 & 7.54E$-$01 & 53.6 \\
 & 20MC & 13441280 & 797508096 & 3.15 & 53.2 & 7.75E$-$01 & 140.6 & 7.53E$-$01 & 55.8 \\
HD102647 & 8MC & 9807616 & 770512384 & 10.49 & 260.3 & 4.64E$-$01 & 160.7 & 4.46E$-$01 & 58.9 \\
 & 28MC & 16618752 & 821404800 & 3.15 & 78.2 & 5.24E$-$01 & 82.8 & 4.81E$-$01 & 53.7 \\
HD102870 & 8MC & 9807360 & 770511744 & 10.49 & 260.3 & 1.13E$-$01 & 42.7 & 6.79E$-$02 & 27.3 \\
HD108903 & 18MC & 13112832 & 791240640 & 3.15 & 53.0 & 8.79E+00 & 184.1 & 1.01E+01 & 117.0 \\
 & 23MC & 15422464 & 806996096 & 3.15 & 150.3 & 8.79E+00 & 100.7 & 1.01E+01 & 200.9 \\
HD110304 & 29MC & 16869376 & 824530176 & 10.49 & 260.1 & 6.18E$-$02 & 18.7 & 7.07E$-$02 & 25.9 \\
HD120933 & 22MC & 15248128 & 804176000 & 3.15 & 77.9 & 6.67E$-$01 & 151.3 & 6.79E$-$01 & 68.8 \\
HD121370 & 29MC & 16836608 & 824525248 & 3.15 & 77.8 & 1.55E$-$01 & 37.1 & 1.57E$-$01 & 38.6 \\
HD123123 & 29MC & 16869632 & 824529280 & 10.49 & 261.0 & 2.86E$-$01 & 88.9 & 2.75E$-$01 & 43.5 \\
HD124897 & 18MC & 13113088 & 791240064 & 3.15 & 53.1 & 7.07E+00 & 181.4 & 8.50E+00 & 132.4 \\
HD131873 & W & 7967744 & 754631104 & 3.15 & 78.3 & 2.05E+00 & 164.3 & 2.09E+00 & 88.2 \\
 & 1MC & 8343040 & 755588096 & 3.15 & 78.2 & 1.95E+00 & 213.6 & 2.03E+00 & 121.3 \\
 & 2MC & 8381952 & 757133952 & 3.15 & 78.4 & 2.03E+00 & 207.7 & 2.12E+00 & 93.9 \\
 & 2MC & 8421376 & 757257280 & 3.15 & 78.4 & 1.92E+00 & 244.1 & 2.06E+00 & 116.1 \\
 & 16MC & 12873728 & 786482624 & 3.15 & 78.5 & 1.94E+00 & 204.6 & 2.11E+00 & 110.8 \\
HD136422 & 23MC & 15421952 & 807195584 & 3.15 & 77.8 & 6.38E$-$01 & 115.3 & 6.37E$-$01 & 63.5 \\
HD138265 & W & 7965696 & 754626368 & 3.15 & 78.2 & 7.66E$-$02 & 12.4 & 7.04E$-$02 & 25.0 \\
 & 5MC & 9192704 & 763724672 & 10.49 & 260.3 & 7.90E$-$02 & 36.1 & 7.36E$-$02 & 36.5 \\
HD140573 & 23MC & 15422208 & 807195072 & 3.15 & 78.3 & 5.19E$-$01 & 146.9 & 5.23E$-$01 & 65.4 \\
HD141477 & 24MC & 15817984 & 809531968 & 3.15 & 78.2 & 6.81E$-$01 & 99.5 & 6.42E$-$01 & 65.9 \\
HD152222 & 5MC & 9192960 & 763699072 & 10.49 & 259.2 & 2.18E$-$02 & 11.3 & 2.47E$-$02 & 19.5 \\
 & 9MC & 9941760 & 772169408 & 10.49 & 511.2 & 3.23E$-$02 & 20.9 & 2.60E$-$02 & 29.4 \\
 & 20MC & 13477632 & 797099968 & 3.15 & 304.4 & 2.00E$-$02 & 8.3 & 2.68E$-$02 & 21.9 \\
HD156283 & 21MC & 13590528 & 801046272 & 3.15 & 78.3 & 6.06E$-$01 & 149.6 & 6.38E$-$01 & 69.8 \\
HD159048 & 17MC & 13078272 & 789150272 & 10.49 & 510.6 & 3.05E$-$02 & 13.9 & 1.90E$-$02 & 19.1 \\
HD159330 & 4MC & 9059584 & 761921344 & 10.49 & 258.8 & 3.63E$-$02 & 16.1 & 3.53E$-$02 & 21.9 \\
 & 20MC & 13477120 & 797222016 & 3.15 & 304.5 & 4.90E$-$02 & 22.2 & 4.86E$-$02 & 38.7 \\
 & 20MC & 13477376 & 797221056 & 10.49 & 511.3 & 4.44E$-$02 & 30.1 & 4.11E$-$02 & 33.6 \\
 & 28MC & 16619520 & 821396800 & 10.49 & 258.7 & 4.54E$-$02 & 18.8 & 4.00E$-$02 & 19.7 \\
HD163588 & R & 7606272 & 753571584 & 3.15 & 71.4 & 2.07E$-$01 & 57.3 & 1.99E$-$01 & 39.8 \\
 & R & 7607040 & 753689280 & 3.15 & 78.0 & 2.10E$-$01 & 31.2 & 1.70E$-$01 & 12.0 \\
 & V & 7795968 & 754091840 & 3.15 & 77.9 & 2.10E$-$01 & 48.4 & 2.08E$-$01 & 42.6 \\
 & W & 7974656 & 754603392 & 3.15 & 78.1 & 2.18E$-$01 & 63.9 & 2.14E$-$01 & 49.3 \\
 & X1 & 7980800 & 753504640 & 3.15 & 77.8 & 2.27E$-$01 & 31.6 & 2.12E$-$01 & 36.5 \\
 & X1 & 7981056 & 753541632 & 3.15 & 78.3 & 2.06E$-$01 & 47.5 & 1.99E$-$01 & 40.8 \\
 & 1MC & 8139008 & 755320896 & 3.15 & 78.0 & 2.16E$-$01 & 69.1 & 2.11E$-$01 & 45.2 \\
 & 1MC & 8140800 & 755397760 & 3.15 & 77.9 & 2.19E$-$01 & 79.0 & 2.12E$-$01 & 45.8 \\
 & 1MC & 8141056 & 755492288 & 3.15 & 78.1 & 2.27E$-$01 & 68.6 & 2.18E$-$01 & 51.7 \\
 & 1MC & 8342272 & 755587008 & 3.15 & 78.1 & 2.09E$-$01 & 78.5 & 2.04E$-$01 & 48.1 \\
 & 1MC & 8783360 & 755756224 & 3.15 & 78.3 & 2.14E$-$01 & 80.6 & 2.12E$-$01 & 50.9 \\
 & 2MC & 8381184 & 757132864 & 3.15 & 78.0 & 2.12E$-$01 & 28.3 & 2.13E$-$01 & 41.8 \\
 & 2MC & 8383232 & 757256256 & 3.15 & 77.9 & 2.15E$-$01 & 57.6 & 2.06E$-$01 & 46.0 \\
 & 3MC & 8809728 & 759525120 & 3.15 & 77.8 & 1.97E$-$01 & 55.7 & 1.92E$-$01 & 37.3 \\
 & 3MC & 8819456 & 759825024 & 3.15 & 77.8 & 2.12E$-$01 & 77.0 & 1.98E$-$01 & 42.2 \\
 & 3MC & 8937728 & 760265344 & 3.15 & 78.0 & 2.01E$-$01 & 34.5 & 2.02E$-$01 & 31.8 \\
 & 4MC & 9067264 & 761709120 & 3.15 & 77.9 & 2.09E$-$01 & 72.2 & 2.00E$-$01 & 42.2 \\
 & 4MC & 9067520 & 762003968 & 3.15 & 78.1 & 2.03E$-$01 & 52.9 & 1.83E$-$01 & 16.9 \\
 & 4MC & 9181696 & 762254336 & 3.15 & 78.0 & 1.72E$-$01 & 22.9 & 1.73E$-$01 & 26.6 \\
 & 5MC & 9191168 & 763690560 & 3.15 & 78.1 & 2.06E$-$01 & 80.4 & 1.99E$-$01 & 43.1 \\
 & 5MC & 9221888 & 764024768 & 3.15 & 78.1 & 2.02E$-$01 & 73.9 & 1.96E$-$01 & 41.9 \\
 & 5MC & 9222656 & 764360512 & 3.15 & 77.9 & 2.10E$-$01 & 70.2 & 2.00E$-$01 & 42.4 \\
 & 6MC & 9617920 & 766301248 & 3.15 & 78.0 & 2.04E$-$01 & 70.7 & 1.98E$-$01 & 41.6 \\
 & 7MC & 9658624 & 767919104 & 3.15 & 77.7 & 2.17E$-$01 & 90.0 & 2.09E$-$01 & 47.0 \\
 & 7MC & 9659136 & 768168000 & 3.15 & 77.9 & 2.08E$-$01 & 64.1 & 2.02E$-$01 & 43.1 \\
 & 8MC & 9802752 & 770187072 & 3.15 & 78.2 & 2.10E$-$01 & 67.7 & 2.01E$-$01 & 41.7 \\
 & 8MC & 9803520 & 770600448 & 3.15 & 78.1 & 2.14E$-$01 & 59.7 & 2.12E$-$01 & 45.2 \\
 & 8MC & 9804288 & 770843904 & 3.15 & 77.9 & 2.07E$-$01 & 62.2 & 2.02E$-$01 & 46.8 \\
 & 9MC & 9937408 & 772025600 & 3.15 & 78.1 & 2.29E$-$01 & 76.9 & 2.17E$-$01 & 46.3 \\
 & 9MC & 9938944 & 772235776 & 3.15 & 77.8 & 2.21E$-$01 & 72.3 & 2.16E$-$01 & 48.3 \\
 & 9MC & 9939712 & 772483136 & 3.15 & 77.9 & 2.23E$-$01 & 57.4 & 2.16E$-$01 & 39.3 \\
 & 10MC & 10088192 & 773705792 & 3.15 & 77.9 & 2.32E$-$01 & 69.7 & 2.19E$-$01 & 46.6 \\
 & 10MC & 10088960 & 773880064 & 3.15 & 77.9 & 2.03E$-$01 & 59.7 & 2.03E$-$01 & 46.4 \\
 & 10MC & 10089728 & 774124928 & 3.15 & 77.9 & 2.32E$-$01 & 69.7 & 2.25E$-$01 & 44.5 \\
 & 11MC & 11780608 & 775520128 & 3.15 & 78.0 & 2.16E$-$01 & 77.6 & 2.18E$-$01 & 49.8 \\
 & 11MC & 11781376 & 775821568 & 3.15 & 77.8 & 2.38E$-$01 & 76.4 & 2.30E$-$01 & 48.7 \\
 & 11MC & 11782144 & 776080512 & 3.15 & 78.2 & 2.34E$-$01 & 90.2 & 2.23E$-$01 & 49.5 \\
 & 11MC & 11782912 & 776280384 & 3.15 & 77.5 & 2.39E$-$01 & 70.4 & 2.11E$-$01 & 34.5 \\
 & 12MC & 11891456 & 777336320 & 3.15 & 78.0 & 2.23E$-$01 & 80.0 & 2.18E$-$01 & 46.4 \\
 & 12MC & 11897344 & 777592128 & 3.15 & 77.8 & 2.22E$-$01 & 69.5 & 2.12E$-$01 & 45.0 \\
 & 12MC & 11898112 & 778057920 & 3.15 & 78.0 & 2.53E$-$01 & 74.5 & 2.26E$-$01 & 49.6 \\
 & 13MC & 12060416 & 779574272 & 3.15 & 78.1 & 2.17E$-$01 & 67.7 & 2.16E$-$01 & 47.0 \\
 & 13MC & 12061184 & 779910912 & 3.15 & 78.1 & 2.33E$-$01 & 67.7 & 2.23E$-$01 & 47.7 \\
 & 13MC & 12061952 & 780221440 & 3.15 & 78.1 & 2.17E$-$01 & 68.6 & 2.17E$-$01 & 49.8 \\
 & 13MC & 12153088 & 780631360 & 3.15 & 77.8 & 2.24E$-$01 & 65.3 & 2.14E$-$01 & 43.1 \\
 & 14MC & 12194816 & 782077056 & 3.15 & 77.9 & 1.99E$-$01 & 61.2 & 2.04E$-$01 & 44.2 \\
 & 14MC & 12195584 & 782377792 & 3.15 & 78.0 & 2.14E$-$01 & 75.5 & 2.13E$-$01 & 47.9 \\
 & 14MC & 12196352 & 782680448 & 3.15 & 77.6 & 2.33E$-$01 & 58.4 & 2.29E$-$01 & 50.1 \\
 & 15MC & 12394752 & 784175424 & 3.15 & 77.7 & 2.14E$-$01 & 67.1 & 2.10E$-$01 & 46.3 \\
 & 15MC & 12395520 & 783803456 & 3.15 & 77.9 & 2.07E$-$01 & 56.8 & 2.04E$-$01 & 43.3 \\
 & 15MC & 12396288 & 784570880 & 3.15 & 78.2 & 2.36E$-$01 & 48.3 & 2.13E$-$01 & 37.2 \\
 & 18MC & 13109504 & 791681152 & 3.15 & 77.9 & 2.22E$-$01 & 83.5 & 2.16E$-$01 & 43.5 \\
 & 18MC & 13110528 & 792003840 & 3.15 & 78.3 & 2.29E$-$01 & 77.2 & 2.16E$-$01 & 46.1 \\
 & 19MC & 13295872 & 793828736 & 3.15 & 77.9 & 2.11E$-$01 & 58.2 & 2.13E$-$01 & 43.6 \\
 & 19MC & 13298688 & 794911040 & 3.15 & 77.9 & 2.17E$-$01 & 60.5 & 2.13E$-$01 & 44.9 \\
 & 19MC & 13299712 & 794452928 & 3.15 & 78.0 & 2.11E$-$01 & 80.9 & 2.11E$-$01 & 44.8 \\
 & 20MC & 13429248 & 796764160 & 3.15 & 77.8 & 2.22E$-$01 & 74.6 & 2.13E$-$01 & 46.1 \\
 & 20MC & 13431808 & 797812480 & 3.15 & 78.1 & 2.22E$-$01 & 61.2 & 2.11E$-$01 & 41.9 \\
 & 20MC & 13432576 & 797325952 & 3.15 & 77.9 & 2.17E$-$01 & 70.4 & 2.10E$-$01 & 45.5 \\
 & 21MC & 13585664 & 801062848 & 3.15 & 77.8 & 2.14E$-$01 & 67.6 & 2.11E$-$01 & 43.4 \\
 & 21MC & 13586432 & 800229760 & 3.15 & 77.7 & 2.07E$-$01 & 77.9 & 2.06E$-$01 & 44.6 \\
 & 21MC & 13587968 & 801073984 & 3.15 & 78.2 & 2.23E$-$01 & 62.2 & 2.18E$-$01 & 46.1 \\
 & 22MC & 15217408 & 803340736 & 3.15 & 77.7 & 2.23E$-$01 & 93.3 & 2.15E$-$01 & 44.5 \\
 & 22MC & 15220736 & 803958208 & 3.15 & 77.6 & 2.22E$-$01 & 58.3 & 2.18E$-$01 & 41.5 \\
 & 22MC & 15221760 & 804509888 & 3.15 & 78.3 & 2.22E$-$01 & 68.3 & 2.19E$-$01 & 47.3 \\
 & 23MC & 15413504 & 806870208 & 3.15 & 78.1 & 2.28E$-$01 & 65.2 & 2.18E$-$01 & 45.9 \\
 & 23MC & 15414528 & 807165184 & 3.15 & 78.3 & 2.38E$-$01 & 60.5 & 2.12E$-$01 & 43.3 \\
 & 23MC & 15415552 & 807541888 & 3.15 & 77.9 & 2.27E$-$01 & 57.5 & 2.01E$-$01 & 30.6 \\
 & 24MC & 15815424 & 809473536 & 3.15 & 78.2 & 2.24E$-$01 & 65.4 & 2.10E$-$01 & 42.5 \\
 & 24MC & 15816448 & 809843584 & 3.15 & 78.1 & 2.21E$-$01 & 67.3 & 2.18E$-$01 & 45.0 \\
 & 24MC & 15817472 & 810383680 & 3.15 & 78.2 & 2.24E$-$01 & 57.1 & 2.05E$-$01 & 44.6 \\
 & 25MC & 15991296 & 811962816 & 3.15 & 78.1 & 2.16E$-$01 & 69.9 & 2.09E$-$01 & 46.2 \\
 & 25MC & 16047872 & 812608960 & 3.15 & 77.5 & 2.21E$-$01 & 60.2 & 2.15E$-$01 & 45.8 \\
 & 25MC & 16048896 & 813301568 & 3.15 & 78.0 & 2.10E$-$01 & 72.6 & 2.04E$-$01 & 43.3 \\
 & 26MC & 16228864 & 815420352 & 3.15 & 77.8 & 2.15E$-$01 & 60.3 & 2.12E$-$01 & 41.8 \\
 & 26MC & 16254464 & 815882816 & 3.15 & 77.9 & 2.05E$-$01 & 58.3 & 2.07E$-$01 & 43.7 \\
 & 26MC & 16255488 & 816317440 & 3.15 & 78.0 & 2.16E$-$01 & 57.9 & 2.13E$-$01 & 43.7 \\
 & 27MC & 16374784 & 817788288 & 3.15 & 77.5 & 2.02E$-$01 & 50.5 & 1.99E$-$01 & 40.3 \\
 & 27MC & 16375552 & 818043200 & 3.15 & 78.1 & 2.21E$-$01 & 56.7 & 2.16E$-$01 & 45.0 \\
 & 29MC & 16834048 & 824364800 & 3.15 & 77.9 & 2.27E$-$01 & 68.5 & 2.14E$-$01 & 44.2 \\
 & 29MC & 16835072 & 824955200 & 3.15 & 78.2 & 2.25E$-$01 & 78.2 & 2.13E$-$01 & 44.1 \\
 & 29MC & 16836096 & 825856512 & 3.15 & 77.8 & 2.12E$-$01 & 68.2 & 2.07E$-$01 & 43.5 \\
HD164058 & 20MC & 13441536 & 797285056 & 3.15 & 53.2 & 1.87E+00 & 224.3 & 2.11E+00 & 101.0 \\
HD166780 & 19MC & 13314816 & 793927168 & 10.49 & 511.0 & 1.96E$-$02 & 7.6 & 1.80E$-$02 & 19.2 \\
HD169916 & 6MC & 9458688 & 766047296 & 10.49 & 259.6 & 3.52E$-$01 & 27.7 & 2.90E$-$01 & 34.8 \\
 & 6MC & 9458944 & 766047872 & 3.15 & 78.1 & 3.45E$-$01 & 25.1 & 2.92E$-$01 & 32.5 \\
HD170693 & W & 7965184 & 754625024 & 3.15 & 77.7 & 8.73E$-$02 & 13.0 & 8.84E$-$02 & 24.7 \\
 & 29MC & 16869888 & 824523456 & 10.49 & 258.7 & 9.39E$-$02 & 18.3 & 9.31E$-$02 & 39.2 \\
HD173398 & 18MC & 13112576 & 791238784 & 10.49 & 510.1 & 4.75E$-$02 & 31.9 & 4.32E$-$02 & 32.6 \\
 & 29MC & 16870144 & 824524224 & 10.49 & 259.5 & 3.69E$-$02 & 19.2 & 4.31E$-$02 & 23.5 \\
HD173511 & 17MC & 12998400 & 789152320 & 10.49 & 510.1 & 1.50E$-$02 & 6.4 & 1.35E$-$02 & 12.3 \\
HD173976 & 6MC & 9459456 & 766034816 & 10.49 & 260.1 & 2.44E$-$02 & 14.2 & 2.36E$-$02 & 19.6 \\
 & 9MC & 9942016 & 772171456 & 10.49 & 509.7 & 2.10E$-$02 & 10.1 & 2.16E$-$02 & 23.0 \\
 & 17MC & 12998144 & 789151360 & 10.49 & 511.3 & 2.10E$-$02 & 13.6 & 2.12E$-$02 & 24.0 \\
 & 20MC & 13478144 & 797808000 & 3.15 & 304.4 & 1.93E$-$02 & 8.3 & 2.20E$-$02 & 21.7 \\
 & 20MC & 13478400 & 797808704 & 10.49 & 511.0 & 1.79E$-$02 & 9.8 & 2.23E$-$02 & 22.8 \\
HD180711 & W & 7966208 & 754627584 & 3.15 & 77.9 & 2.95E$-$01 & 45.7 & 2.79E$-$01 & 43.4 \\
 & 4MC & 9068032 & 761710080 & 3.15 & 78.1 & 2.69E$-$01 & 68.3 & 2.61E$-$01 & 39.5 \\
 & 4MC & 9068288 & 762004928 & 3.15 & 78.0 & 2.70E$-$01 & 70.5 & 2.52E$-$01 & 30.8 \\
 & 4MC & 9181952 & 762260928 & 3.15 & 77.7 & 2.35E$-$01 & 29.3 & 2.39E$-$01 & 28.1 \\
 & 5MC & 9191424 & 763691456 & 3.15 & 78.2 & 2.84E$-$01 & 84.1 & 2.73E$-$01 & 49.5 \\
 & 5MC & 9222144 & 764025664 & 3.15 & 78.1 & 2.76E$-$01 & 98.5 & 2.48E$-$01 & 24.4 \\
 & 8MC & 9803008 & 770188032 & 3.15 & 78.4 & 2.68E$-$01 & 68.6 & 2.63E$-$01 & 46.9 \\
 & 8MC & 9803776 & 770601344 & 3.15 & 78.0 & 2.69E$-$01 & 80.1 & 2.67E$-$01 & 49.7 \\
 & 8MC & 9804544 & 770844800 & 3.15 & 78.2 & 2.59E$-$01 & 85.1 & 2.52E$-$01 & 43.6 \\
 & 9MC & 9937664 & 772026496 & 3.15 & 78.4 & 2.98E$-$01 & 95.9 & 2.82E$-$01 & 53.1 \\
 & 9MC & 9939200 & 772236672 & 3.15 & 77.7 & 2.72E$-$01 & 76.1 & 2.73E$-$01 & 48.3 \\
 & 9MC & 9939968 & 772484032 & 3.15 & 78.1 & 3.01E$-$01 & 71.7 & 2.90E$-$01 & 50.6 \\
 & 10MC & 10088448 & 773706752 & 3.15 & 77.8 & 2.89E$-$01 & 72.1 & 2.77E$-$01 & 45.2 \\
 & 10MC & 10089216 & 773881024 & 3.15 & 77.9 & 2.78E$-$01 & 85.2 & 2.74E$-$01 & 49.5 \\
 & 10MC & 10089984 & 774125888 & 3.15 & 78.1 & 3.03E$-$01 & 89.8 & 2.90E$-$01 & 49.5 \\
 & 11MC & 11780864 & 775521024 & 3.15 & 78.1 & 2.89E$-$01 & 82.2 & 2.80E$-$01 & 50.4 \\
 & 11MC & 11781632 & 775822464 & 3.15 & 78.0 & 2.95E$-$01 & 80.8 & 2.76E$-$01 & 51.5 \\
 & 11MC & 11782400 & 776081472 & 3.15 & 77.8 & 3.02E$-$01 & 84.1 & 2.90E$-$01 & 50.5 \\
 & 11MC & 11783168 & 776281344 & 3.15 & 78.3 & 2.80E$-$01 & 82.7 & 2.82E$-$01 & 51.5 \\
 & 12MC & 11891712 & 777337280 & 3.15 & 78.0 & 2.90E$-$01 & 84.9 & 2.70E$-$01 & 48.5 \\
 & 12MC & 11897600 & 777593088 & 3.15 & 77.7 & 2.80E$-$01 & 77.5 & 2.79E$-$01 & 45.6 \\
 & 12MC & 11898368 & 778058816 & 3.15 & 77.9 & 2.83E$-$01 & 63.4 & 2.76E$-$01 & 45.2 \\
 & 13MC & 12060672 & 779575168 & 3.15 & 78.1 & 3.13E$-$01 & 86.8 & 2.74E$-$01 & 38.6 \\
 & 13MC & 12061440 & 779911808 & 3.15 & 78.3 & 2.99E$-$01 & 90.1 & 2.91E$-$01 & 51.0 \\
 & 13MC & 12062208 & 780222336 & 3.15 & 78.3 & 3.02E$-$01 & 104.3 & 2.98E$-$01 & 53.2 \\
 & 13MC & 12153344 & 780632320 & 3.15 & 77.8 & 2.95E$-$01 & 82.4 & 2.86E$-$01 & 48.9 \\
 & 14MC & 12195072 & 782077952 & 3.15 & 77.9 & 2.78E$-$01 & 83.5 & 2.70E$-$01 & 49.0 \\
 & 14MC & 12195840 & 782378752 & 3.15 & 77.7 & 2.80E$-$01 & 79.2 & 2.79E$-$01 & 51.4 \\
 & 14MC & 12196608 & 782681408 & 3.15 & 78.3 & 3.09E$-$01 & 86.2 & 3.07E$-$01 & 60.3 \\
 & 15MC & 12395008 & 784176384 & 3.15 & 78.3 & 2.85E$-$01 & 77.2 & 2.83E$-$01 & 52.8 \\
 & 15MC & 12395776 & 783804352 & 3.15 & 78.0 & 2.76E$-$01 & 71.6 & 2.69E$-$01 & 47.5 \\
 & 15MC & 12396544 & 784571840 & 3.15 & 78.1 & 2.96E$-$01 & 61.9 & 2.85E$-$01 & 46.5 \\
 & 16MC & 12871680 & 786141952 & 3.15 & 78.0 & 2.90E$-$01 & 85.8 & 2.81E$-$01 & 49.0 \\
 & 16MC & 12884480 & 786574272 & 3.15 & 78.4 & 3.03E$-$01 & 90.2 & 2.94E$-$01 & 50.5 \\
 & 16MC & 12884992 & 786900864 & 3.15 & 77.9 & 2.81E$-$01 & 74.9 & 2.72E$-$01 & 47.0 \\
 & 17MC & 12997376 & 788129216 & 3.15 & 78.0 & 2.92E$-$01 & 107.2 & 2.82E$-$01 & 52.6 \\
 & 17MC & 13001728 & 788494656 & 3.15 & 78.1 & 2.77E$-$01 & 67.9 & 2.78E$-$01 & 48.0 \\
 & 17MC & 13072896 & 789156032 & 3.15 & 77.9 & 2.99E$-$01 & 82.4 & 2.83E$-$01 & 45.2 \\
 & 18MC & 13109248 & 790894272 & 3.15 & 78.2 & 2.75E$-$01 & 65.2 & 2.72E$-$01 & 46.3 \\
 & 18MC & 13110272 & 792002816 & 3.15 & 77.9 & 2.77E$-$01 & 91.4 & 2.70E$-$01 & 47.4 \\
 & 18MC & 13111296 & 791342528 & 3.15 & 77.8 & 2.80E$-$01 & 89.0 & 2.74E$-$01 & 48.8 \\
 & 19MC & 13295616 & 793827456 & 3.15 & 77.9 & 2.88E$-$01 & 99.6 & 2.66E$-$01 & 48.5 \\
 & 19MC & 13298432 & 794454208 & 3.15 & 78.3 & 2.86E$-$01 & 92.2 & 2.81E$-$01 & 51.6 \\
 & 19MC & 13299456 & 794910016 & 3.15 & 78.3 & 2.82E$-$01 & 99.2 & 2.78E$-$01 & 53.1 \\
 & 22MC & 15217664 & 803341952 & 3.15 & 77.9 & 2.94E$-$01 & 77.2 & 2.78E$-$01 & 48.3 \\
 & 22MC & 15220992 & 803956992 & 3.15 & 77.8 & 2.87E$-$01 & 86.7 & 2.82E$-$01 & 47.4 \\
 & 22MC & 15222016 & 804511168 & 3.15 & 78.0 & 2.91E$-$01 & 76.4 & 2.86E$-$01 & 51.7 \\
 & 23MC & 15413760 & 806871168 & 3.15 & 77.2 & 2.75E$-$01 & 71.4 & 2.71E$-$01 & 46.3 \\
 & 23MC & 15414784 & 807166400 & 3.15 & 78.3 & 2.72E$-$01 & 86.6 & 2.66E$-$01 & 46.3 \\
 & 23MC & 15415808 & 807540672 & 3.15 & 78.0 & 2.76E$-$01 & 70.5 & 2.77E$-$01 & 50.4 \\
 & 24MC & 15815680 & 809474496 & 3.15 & 78.2 & 2.81E$-$01 & 81.6 & 2.75E$-$01 & 49.1 \\
 & 24MC & 15816704 & 809844544 & 3.15 & 78.3 & 2.85E$-$01 & 84.5 & 2.82E$-$01 & 49.7 \\
 & 24MC & 15817728 & 810384640 & 3.15 & 77.9 & 2.72E$-$01 & 81.0 & 2.71E$-$01 & 45.8 \\
 & 25MC & 15991552 & 811963136 & 3.15 & 78.0 & 2.85E$-$01 & 83.1 & 2.69E$-$01 & 45.0 \\
 & 25MC & 16048128 & 812608576 & 3.15 & 78.1 & 2.88E$-$01 & 83.2 & 2.79E$-$01 & 49.4 \\
 & 25MC & 16049152 & 813301952 & 3.15 & 77.6 & 2.75E$-$01 & 82.0 & 2.55E$-$01 & 47.0 \\
 & 26MC & 16229120 & 815420672 & 3.15 & 78.0 & 2.91E$-$01 & 90.5 & 2.76E$-$01 & 48.6 \\
 & 26MC & 16254720 & 815883392 & 3.15 & 77.7 & 2.99E$-$01 & 78.0 & 2.72E$-$01 & 48.7 \\
 & 26MC & 16255744 & 816317824 & 3.15 & 77.5 & 2.73E$-$01 & 75.4 & 2.70E$-$01 & 48.8 \\
 & 27MC & 16375040 & 817789248 & 3.15 & 77.6 & 2.86E$-$01 & 60.7 & 2.78E$-$01 & 43.6 \\
 & 27MC & 16375808 & 818044224 & 3.15 & 77.9 & 2.88E$-$01 & 92.1 & 2.82E$-$01 & 48.4 \\
 & 27MC & 16377344 & 818481664 & 3.15 & 77.8 & 3.07E$-$01 & 74.5 & 2.76E$-$01 & 47.6 \\
 & 28MC & 16603136 & 820974720 & 3.15 & 78.0 & 2.80E$-$01 & 85.3 & 2.67E$-$01 & 46.9 \\
 & 28MC & 16603904 & 821395200 & 3.15 & 78.0 & 2.93E$-$01 & 90.5 & 2.72E$-$01 & 48.4 \\
 & 28MC & 16604672 & 821607616 & 3.15 & 77.8 & 2.87E$-$01 & 96.1 & 2.77E$-$01 & 48.0 \\
 & 29MC & 16834304 & 824363584 & 3.15 & 77.8 & 2.94E$-$01 & 85.6 & 2.81E$-$01 & 46.3 \\
 & 29MC & 16835328 & 824956160 & 3.15 & 77.9 & 2.86E$-$01 & 91.6 & 2.74E$-$01 & 47.0 \\
 & 29MC & 16836352 & 825855040 & 3.15 & 77.8 & 2.82E$-$01 & 92.7 & 2.72E$-$01 & 48.4 \\
HD183439 & 7MC & 9662208 & 767993792 & 3.15 & 78.1 & 3.32E$-$01 & 77.4 & 3.43E$-$01 & 46.3 \\
HD197989 & 21MC & 13590784 & 801047296 & 3.15 & 78.0 & 5.02E$-$01 & 105.7 & 4.95E$-$01 & 52.7 \\
HD198542 & 21MC & 13591040 & 801047872 & 3.15 & 78.2 & 5.25E$-$01 & 114.7 & 5.06E$-$01 & 55.2 \\
HD209952 & R & 7602176 & 753650688 & 10.49 & 259.3 & 8.28E$-$02 & 22.5 & 7.11E$-$02 & 29.8 \\
 & X1 & 7979008 & 753516608 & 3.15 & 77.5 & 9.69E$-$02 & 13.4 & 7.77E$-$02 & 20.4 \\
 & 7MC & 9661696 & 768552960 & 10.49 & 259.1 & 7.19E$-$02 & 30.9 & 7.39E$-$02 & 34.7 \\
 & 21MC & 13642240 & 800777728 & 3.15 & 153.4 & 7.70E$-$02 & 32.1 & 6.89E$-$02 & 39.0 \\
HD213310 & W & 7966976 & 754629376 & 3.15 & 78.1 & 6.72E$-$01 & 79.6 & 6.74E$-$01 & 54.6 \\
 & 22MC & 15248384 & 804442624 & 3.15 & 78.2 & 6.60E$-$01 & 155.1 & 6.78E$-$01 & 62.0 \\
HD216131 & W & 7965952 & 754627008 & 3.15 & 77.6 & 2.30E$-$01 & 42.2 & 2.21E$-$01 & 40.6 \\
HD217906 & 11MC & 11784192 & 775641728 & 3.15 & 78.4 & 4.55E+00 & 235.2 & 5.42E+00 & 132.5 \\
\enddata
\tablenotetext{a}{The letter (for In-Orbit-Checkout) or number of the
MIPS Campaign (MC) in which the observations were taken.}
\tablenotetext{b}{The time the exposure started is the number of
seconds from 1/1/1980.}
\tablenotetext{c}{The total exposure time is computed as the average
exposure time per pixel in the object aperture to account for exposure
time lost to cosmic ray rejection.}
\end{deluxetable}

\clearpage

\begin{deluxetable}{lrrrrrrrrr}
\tablewidth{0pt}
\tablecaption{Fine-scale Measurements \label{tab_meas_nf}}
\tablehead{\colhead{Name} & \colhead{Campaign\tablenotemark{a}} &
    \colhead{AOR \#} & \colhead{Time\tablenotemark{b}} & 
    \colhead{ExpTime} & \colhead{TotTime\tablenotemark{c}} &
    \colhead{Ap. Flux} & \colhead{S/N} & \colhead{PSF Flux} & \colhead{S/N} \\
 & & & \colhead{[s]} & \colhead{[s]} & \colhead{[s]} & 
    \colhead{[MIPS70F]} & & \colhead{[MIPS70F]} & }
\startdata
HD045348 & 6MC & 9456384 & 765978944 & 10.49 & 187.1 & 1.66E+00 & 139.0 & 1.56E+00 & 111.0 \\
HD048915 & 6MC & 9456896 & 765977024 & 10.49 & 186.3 & 1.52E+00 & 119.2 & 1.47E+00 & 118.2 \\
HD071129 & 29MC & 16863744 & 825634368 & 10.49 & 124.3 & 2.90E+00 & 208.3 & 2.68E+00 & 102.0 \\
HD080493 & 21MC & 13635072 & 800352640 & 10.49 & 248.7 & 9.89E$-$01 & 148.9 & 9.03E$-$01 & 125.0 \\
HD082668 & 7MC & 9653248 & 768348480 & 10.49 & 248.4 & 9.53E$-$01 & 67.9 & 8.75E$-$01 & 126.5 \\
HD100029 & 5MC & 9189888 & 763723136 & 10.49 & 186.7 & 6.47E$-$01 & 104.2 & 6.20E$-$01 & 100.8 \\
HD108903 & 18MC & 13113600 & 791361728 & 10.49 & 251.2 & 1.09E+01 & 282.4 & 9.83E+00 & 163.8 \\
HD131873 & 1MC & 8343296 & 755588352 & 3.15 & 56.0 & 1.80E+00 & 106.0 & 1.76E+00 & 110.6 \\
 & 2MC & 8382208 & 757134208 & 3.15 & 55.7 & 1.76E+00 & 88.0 & 1.63E+00 & 100.7 \\
 & 2MC & 8422400 & 757257536 & 3.15 & 55.7 & 1.81E+00 & 92.2 & 1.71E+00 & 100.0 \\
 & 5MC & 9190400 & 763697536 & 10.49 & 186.9 & 1.76E+00 & 201.6 & 1.62E+00 & 93.8 \\
HD163588 & R & 7606528 & 753571776 & 3.15 & 55.8 & 2.29E$-$01 & 15.7 & 1.78E$-$01 & 31.7 \\
 & R & 7607296 & 753689536 & 3.15 & 55.7 & 1.71E$-$01 & 10.1 & 1.67E$-$01 & 32.8 \\
 & V & 7796224 & 754091456 & 3.15 & 55.3 & 2.21E$-$01 & 12.2 & 1.94E$-$01 & 34.7 \\
 & W & 7974912 & 754603648 & 3.15 & 55.6 & 2.20E$-$01 & 13.7 & 2.00E$-$01 & 32.9 \\
 & 1MC & 8141568 & 755321152 & 3.15 & 55.5 & 2.25E$-$01 & 16.3 & 1.81E$-$01 & 38.2 \\
 & 1MC & 8141824 & 755398016 & 3.15 & 55.7 & 1.66E$-$01 & 9.7 & 1.99E$-$01 & 34.1 \\
 & 1MC & 8142080 & 755492480 & 3.15 & 55.4 & 2.01E$-$01 & 14.8 & 1.94E$-$01 & 35.3 \\
 & 1MC & 8342528 & 755587264 & 3.15 & 55.6 & 2.44E$-$01 & 15.3 & 1.96E$-$01 & 35.6 \\
 & 1MC & 8783616 & 755756480 & 3.15 & 55.6 & 1.67E$-$01 & 12.5 & 1.87E$-$01 & 45.0 \\
 & 2MC & 8381440 & 757133120 & 3.15 & 55.7 & 1.85E$-$01 & 9.0 & 1.65E$-$01 & 28.7 \\
 & 2MC & 8384256 & 757256448 & 3.15 & 55.6 & 2.43E$-$01 & 15.2 & 1.84E$-$01 & 37.1 \\
 & 26MC & 16276736 & 815937216 & 10.49 & 562.2 & 1.99E$-$01 & 43.0 & 1.77E$-$01 & 70.3 \\
HD217906 & 16MC & 12868608 & 786224896 & 10.49 & 249.9 & 5.14E+00 & 369.8 & 4.70E+00 & 136.8 \\
\enddata
\tablenotetext{a}{The letter (for In-Orbit-Checkout) or number of the
MIPS Campaign (MC) in which the observations were taken.}
\tablenotetext{b}{The time the exposure started is the number of
seconds from 1/1/1980.}
\tablenotetext{c}{The total exposure time is computed as the average
exposure time per pixel in the object aperture to account for exposure
time lost to cosmic ray rejection.}
\end{deluxetable}

\clearpage

\begin{deluxetable}{llrrrrrrrrrrrr}
\tablewidth{0pt}
\tablecaption{Coarse-Scale Calibration Factors \label{tab_ave_wf}}
\tablehead{ & & \multicolumn{2}{c}{Predicted} &
    \multicolumn{2}{c}{Background} & \multicolumn{4}{c}{Average
    Measured\tablenotemark{b}} & \multicolumn{4}{c}{Calibration Factor} \\
    \colhead{Name\tablenotemark{a}} & \colhead{SpType} & \colhead{Flux} &
    \colhead{Unc.} &\colhead{SB} & \colhead{Unc} & 
    \colhead{Ap. Flux} & \colhead{S/N} & 
    \colhead{PSF Flux} & \colhead{S/N} & 
    \colhead{Ap.} & \colhead{Unc.} & \colhead{PSF} & \colhead{Unc.}  \\
 & & \multicolumn{2}{c}{[mJy]} & \multicolumn{2}{c}{[MJy sr$^{-1}$]} & 
   \colhead{[MIPS70]} & &
   \colhead{[MIPS70]} & & \multicolumn{4}{c}{[MJy sr$^{-1}$ MIPS70$^{-1}$]}}
\startdata
HD002151& G2IV & 255.6 & 8.9 & 5.3 & 0.6 & 1.53E$-$01 & 54.2 & 1.59E$-$01 & 68.8 & 734.8 & 29.0 & 706.0 & 26.7 \\
HD002261& K0III & 1138.0 & 59.6 & 6.3 & 0.5 & 7.15E$-$01 & 49.7 & 6.88E$-$01 & 58.3 & 697.5 & 39.2 & 725.5 & 40.0 \\
HD003712& K0IIIa & 1180.0 & 47.6 & 10.9 & 0.8 & 7.11E$-$01 & 55.5 & 7.37E$-$01 & 70.2 & 727.9 & 32.2 & 702.3 & 30.1 \\
HD004128& K0III & 1205.0 & 40.1 & 10.1 & 1.1 & 6.98E$-$01 & 52.7 & 6.90E$-$01 & 63.6 & 757.5 & 29.0 & 765.7 & 28.2 \\
HD006860& M0III & 5334.0 & 202.2 & 8.5 & 1.3 & 3.04E+00 & 73.7 & 3.35E+00 & 98.8 & 768.4 & 30.9 & 699.2 & 27.4 \\
HD009053& M0IIIa & 1504.0 & 64.9 & 5.7 & 0.3 & 8.85E$-$01 & 62.2 & 9.10E$-$01 & 78.0 & 744.9 & 34.3 & 724.5 & 32.6 \\
HD009927& K3III & 472.2 & 15.8 & 8.9 & 1.0 & 2.95E$-$01 & 57.7 & 2.84E$-$01 & 67.8 & 701.6 & 26.4 & 728.4 & 26.6 \\
HD012533& K3IIb & 1985.0 & 67.2 & 8.8 & 1.3 & 1.24E+00 & 63.8 & 1.30E+00 & 81.5 & 699.5 & 26.1 & 667.9 & 24.1 \\
HD012929& K2III & 1734.0 & 57.8 & 13.7 & 2.7 & 1.04E+00 & 56.0 & 1.08E+00 & 70.7 & 731.5 & 27.7 & 706.1 & 25.6 \\
HD015008& A1/2V & 22.0 & 0.8 & 4.7 & 0.6 & 1.79E$-$02 & 26.3 & 1.25E$-$02 & 22.4 & 539.0 & 28.3 & 769.9 & 44.1 \\
HD018884& M1.5III & 4645.0 & 160.1 & 14.4 & 2.2 & 2.60E+00 & 80.0 & 2.88E+00 & 108.2 & 784.6 & 28.8 & 708.1 & 25.3 \\
HD020902& F5I & 510.7 & 19.2 & 15.2 & 1.4 & 3.13E$-$01 & 26.3 & 3.02E$-$01 & 30.9 & 714.5 & 38.2 & 741.9 & 36.8 \\
HD024512& M2III & 2421.0 & 94.0 & 6.1 & 0.6 & 1.41E+00 & 102.1 & 1.47E+00 & 130.3 & 754.7 & 30.2 & 721.4 & 28.6 \\
HD025025& M1IIIb & 2320.0 & 82.5 & 8.0 & 0.8 & 1.48E+00 & 67.3 & 1.55E+00 & 86.1 & 687.5 & 26.5 & 655.5 & 24.5 \\
HD029139& K5III & 12840.0 & 451.2 & 23.7 & 3.1 & 5.96E+00 & 75.0 & 7.13E+00 & 109.5 & 944.8 & 35.5 & 790.1 & 28.7 \\
HD031398*& K3II & 1560.0 & 65.8 & 26.6 & 2.9 & 1.11E+00 & 60.4 & 1.20E+00 & 79.3 & 615.5 & 27.9 & 572.1 & 25.2 \\
HD032887& K4III & 1159.0 & 40.5 & 6.1 & 0.8 & 7.31E$-$01 & 50.7 & 7.25E$-$01 & 61.3 & 695.0 & 27.9 & 700.9 & 27.0 \\
HD034029& G5IIIe+ & 5095.0 & 201.8 & 19.4 & 1.8 & 2.72E+00 & 127.2 & 2.98E+00 & 170.0 & 821.1 & 33.2 & 750.0 & 30.0 \\
HD035666*& K3III & 26.9 & 1.1 & 6.2 & 0.5 & 1.77E$-$02 & 19.5 & 1.76E$-$02 & 23.6 & 665.4 & 43.1 & 671.5 & 38.9 \\
HD036167*& K5III & 417.1 & 26.5 & 17.5 & 1.6 & 2.56E$-$01 & 45.9 & 2.36E$-$01 & 51.7 & 714.5 & 47.9 & 774.1 & 51.3 \\
HD039425& K2III & 587.7 & 19.2 & 5.2 & 0.7 & 3.52E$-$01 & 14.7 & 2.89E$-$01 & 14.7 & 732.5 & 55.4 & 891.2 & 67.2 \\
HD039608& K5III & 29.7 & 1.2 & 5.0 & 0.5 & 2.66E$-$02 & 32.7 & 2.14E$-$02 & 32.0 & 488.2 & 24.7 & 608.5 & 31.0 \\
HD042701*& K3III & 33.8 & 1.0 & 5.1 & 0.6 & 6.10E$-$02 & 54.8 & 3.18E$-$02 & 34.8 & 242.7 & 8.1 & 465.5 & 18.7 \\
HD045348& F0II & 3085.0 & 67.1 & 5.6 & 0.5 & 1.75E+00 & 89.5 & 1.84E+00 & 114.9 & 772.2 & 18.9 & 733.5 & 17.2 \\
HD048915& A0V & 2900.0 & 354.5 & 14.8 & 1.0 & 1.60E+00 & 43.8 & 1.52E+00 & 50.8 & 793.3 & 98.6 & 834.1 & 103.3 \\
HD050310& K1III & 706.0 & 24.7 & 5.8 & 0.6 & 4.37E$-$01 & 67.4 & 4.26E$-$01 & 80.2 & 708.0 & 26.9 & 725.9 & 27.0 \\
HD051799& M1III & 593.1 & 20.0 & 5.4 & 0.6 & 3.98E$-$01 & 57.8 & 3.84E$-$01 & 68.2 & 654.2 & 24.8 & 676.7 & 24.9 \\
HD053501& K3III & 143.3 & 4.5 & 6.0 & 0.6 & 9.95E$-$02 & 43.6 & 8.30E$-$02 & 44.4 & 631.8 & 24.5 & 757.3 & 29.1 \\
HD056855& K3Ib & 2544.0 & 81.3 & 8.7 & 0.7 & 1.61E+00 & 63.4 & 1.77E+00 & 85.4 & 694.3 & 24.7 & 629.5 & 21.4 \\
HD059717& K5III & 1423.0 & 49.0 & 8.2 & 0.6 & 8.89E$-$01 & 60.7 & 9.23E$-$01 & 76.9 & 702.3 & 26.8 & 675.9 & 24.9 \\
HD060522& M0III & 746.5 & 25.5 & 15.2 & 3.2 & 4.54E$-$01 & 34.4 & 4.56E$-$01 & 42.1 & 720.5 & 32.3 & 718.1 & 29.9 \\
HD062509& K0IIIb & 2607.0 & 79.1 & 14.7 & 3.1 & 1.56E+00 & 73.2 & 1.64E+00 & 93.7 & 732.1 & 24.4 & 697.0 & 22.4 \\
HD071129& K3III+ & 5153.0 & 168.7 & 7.5 & 0.6 & 2.85E+00 & 162.6 & 3.15E+00 & 218.9 & 792.9 & 26.4 & 718.4 & 23.7 \\
HD080007& A2IV & 199.9 & 6.1 & 5.6 & 0.6 & 1.38E$-$01 & 47.2 & 1.26E$-$01 & 52.7 & 635.0 & 23.5 & 693.8 & 24.8 \\
HD080493& K7III & 1686.0 & 60.0 & 10.8 & 1.9 & 1.02E+00 & 86.4 & 1.07E+00 & 110.0 & 721.6 & 27.0 & 691.1 & 25.4 \\
HD081797& K3II-III & 2887.0 & 101.1 & 9.2 & 1.7 & 1.76E+00 & 74.3 & 1.86E+00 & 95.9 & 720.1 & 27.0 & 680.8 & 24.9 \\
HD082308& K5III & 542.7 & 20.3 & 14.5 & 2.9 & 3.28E$-$01 & 44.3 & 3.21E$-$01 & 52.9 & 725.3 & 31.6 & 742.0 & 31.1 \\
HD082668& K5III & 1424.0 & 204.5 & 25.7 & 0.5 & 9.36E$-$01 & 69.9 & 9.29E$-$01 & 84.6 & 667.1 & 96.3 & 672.4 & 96.9 \\
HD087901& B7 & 196.0 & 5.6 & 15.3 & 3.3 & 1.04E$-$01 & 41.6 & 1.08E$-$01 & 52.8 & 824.3 & 30.7 & 793.0 & 27.1 \\
HD089388& K3IIa & 820.1 & 187.9 & 14.1 & 0.6 & 6.04E$-$01 & 39.8 & 6.11E$-$01 & 49.0 & 595.4 & 137.2 & 589.0 & 135.5 \\
HD089484& K1IIIb & 1744.0 & 52.1 & 14.1 & 2.7 & 1.03E+00 & 100.3 & 1.06E+00 & 125.8 & 741.7 & 23.4 & 721.4 & 22.3 \\
HD089758& M0III & 2192.0 & 78.5 & 8.1 & 1.1 & 1.37E+00 & 57.8 & 1.37E+00 & 70.8 & 702.8 & 28.0 & 699.4 & 26.9 \\
HD092305& M0III & 746.3 & 29.8 & 9.6 & 0.6 & 4.81E$-$01 & 68.0 & 4.85E$-$01 & 83.6 & 680.2 & 28.9 & 675.0 & 28.1 \\
HD093813& K0/K1III & 699.0 & 23.7 & 9.3 & 1.5 & 4.57E$-$01 & 47.1 & 4.70E$-$01 & 59.2 & 671.4 & 26.9 & 652.4 & 24.7 \\
HD095689& K0Iab & 1694.0 & 53.2 & 5.4 & 0.6 & 1.07E+00 & 72.7 & 1.10E+00 & 91.3 & 694.4 & 23.8 & 674.5 & 22.4 \\
HD096833& K1III & 626.2 & 20.1 & 7.0 & 0.7 & 3.89E$-$01 & 52.5 & 3.64E$-$01 & 60.0 & 706.2 & 26.3 & 753.6 & 27.2 \\
HD100029& M0III & 1124.0 & 38.2 & 4.9 & 0.6 & 7.76E$-$01 & 65.3 & 7.53E$-$01 & 77.4 & 635.5 & 23.7 & 654.2 & 23.8 \\
HD102647*& A3V & 141.4 & 4.5 & 13.2 & 2.0 & 4.89E$-$01 & 69.2 & 4.61E$-$01 & 79.6 & 126.8 & 4.4 & 134.5 & 4.6 \\
HD102870& F9V & 108.1 & 7.3 & 15.2 & 3.3 & 1.13E$-$01 & 37.3 & 6.79E$-$02 & 27.3 & 418.0 & 30.2 & 698.1 & 53.4 \\
HD108903& M3.5III & 17000.0 & 1766.0 & 18.4 & 0.8 & 8.79E+00 & 165.2 & 1.01E+01 & 232.5 & 848.5 & 88.3 & 735.7 & 76.5 \\
HD110304& A1IV & 119.8 & 3.2 & 7.8 & 0.8 & 6.18E$-$02 & 18.6 & 7.07E$-$02 & 25.9 & 850.7 & 51.1 & 742.7 & 34.8 \\
HD120933*& K5III & 834.4 & 30.6 & 6.2 & 0.2 & 6.67E$-$01 & 55.4 & 6.79E$-$01 & 68.8 & 548.5 & 22.4 & 539.0 & 21.3 \\
HD121370& G0IV & 260.9 & 10.2 & 8.3 & 0.7 & 1.55E$-$01 & 31.2 & 1.57E$-$01 & 38.6 & 740.1 & 37.5 & 730.5 & 34.3 \\
HD123123& K2III & 490.9 & 16.4 & 12.2 & 2.2 & 2.86E$-$01 & 37.1 & 2.75E$-$01 & 43.5 & 751.8 & 32.2 & 782.6 & 31.7 \\
HD124897& K1.5III & 14340.0 & 778.8 & 7.9 & 0.6 & 7.07E+00 & 90.2 & 8.50E+00 & 132.4 & 889.9 & 49.3 & 740.0 & 40.6 \\
HD131873& K4III & 3363.0 & 123.7 & 4.4 & 0.5 & 1.97E+00 & 185.6 & 2.07E+00 & 238.9 & 749.6 & 27.9 & 710.7 & 26.3 \\
HD136422& K5III & 1042.0 & 41.0 & 13.6 & 2.1 & 6.38E$-$01 & 52.2 & 6.37E$-$01 & 63.5 & 715.8 & 31.3 & 717.4 & 30.4 \\
HD138265& K5III & 111.4 & 3.8 & 4.4 & 0.4 & 7.82E$-$02 & 39.1 & 7.25E$-$02 & 44.2 & 625.0 & 26.6 & 673.8 & 27.5 \\
HD140573& K2IIIb & 883.8 & 28.3 & 9.1 & 1.2 & 5.19E$-$01 & 53.2 & 5.23E$-$01 & 65.4 & 746.9 & 27.7 & 740.5 & 26.3 \\
HD141477*& M0.5III & 921.5 & 32.4 & 7.1 & 0.7 & 6.81E$-$01 & 57.3 & 6.42E$-$01 & 65.9 & 593.5 & 23.3 & 629.5 & 24.1 \\
HD152222& K2III & 37.9 & 1.8 & 4.5 & 0.5 & 2.65E$-$02 & 34.8 & 2.59E$-$02 & 41.5 & 626.7 & 34.2 & 641.8 & 33.6 \\
HD156283& K3Iab & 940.4 & 30.8 & 5.1 & 0.6 & 6.06E$-$01 & 54.3 & 6.38E$-$01 & 69.8 & 680.9 & 25.6 & 646.0 & 23.1 \\
HD159048& K0III & 26.9 & 0.9 & 4.7 & 0.5 & 3.05E$-$02 & 25.1 & 1.90E$-$02 & 19.1 & 387.0 & 19.9 & 622.5 & 38.4 \\
HD159330& K2III & 61.7 & 2.0 & 4.6 & 0.4 & 4.44E$-$02 & 50.6 & 4.23E$-$02 & 58.8 & 608.6 & 22.9 & 639.5 & 23.2 \\
HD163588& K2III & 354.4 & 9.9 & 4.7 & 0.4 & 2.18E$-$01 & 335.0 & 2.11E$-$01 & 395.7 & 714.3 & 20.1 & 737.8 & 20.7 \\
HD164058& K5III & 3315.0 & 99.7 & 4.9 & 0.5 & 1.87E+00 & 73.6 & 2.11E+00 & 101.0 & 775.7 & 25.6 & 690.2 & 21.8 \\
HD166780*& K4.5III & 23.8 & 1.0 & 4.9 & 0.4 & 1.96E$-$02 & 17.1 & 1.80E$-$02 & 19.2 & 532.7 & 37.9 & 580.9 & 38.4 \\
HD169916*& K1IIIb & 675.8 & 27.5 & 22.3 & 3.3 & 3.49E$-$01 & 46.8 & 2.91E$-$01 & 47.6 & 849.5 & 39.1 & 1019.1 & 46.7 \\
HD170693& K1.5III & 147.7 & 4.4 & 4.8 & 0.5 & 9.19E$-$02 & 38.1 & 9.16E$-$02 & 46.4 & 705.0 & 28.0 & 706.7 & 25.9 \\
HD173398& K0III & 23.9 & 0.9 & 4.8 & 0.5 & 4.39E$-$02 & 33.4 & 4.32E$-$02 & 40.1 & 238.6 & 11.6 & 242.3 & 11.1 \\
HD173511*& K5III & 24.5 & 0.9 & 4.8 & 0.5 & 1.50E$-$02 & 11.2 & 1.35E$-$02 & 12.3 & 713.9 & 68.3 & 792.5 & 70.0 \\
HD173976*& K5III & 35.8 & 1.2 & 4.6 & 0.5 & 2.05E$-$02 & 37.9 & 2.20E$-$02 & 49.7 & 765.6 & 32.9 & 712.6 & 28.1 \\
HD180711& G9III & 447.4 & 11.9 & 4.8 & 0.6 & 2.86E$-$01 & 333.7 & 2.76E$-$01 & 393.8 & 686.6 & 18.3 & 709.8 & 18.9 \\
HD183439& M0III & 575.2 & 24.9 & 20.6 & 0.9 & 3.32E$-$01 & 36.7 & 3.43E$-$01 & 46.3 & 760.4 & 38.9 & 735.2 & 35.5 \\
HD197989& K0III & 850.9 & 34.3 & 11.2 & 0.8 & 5.02E$-$01 & 43.8 & 4.95E$-$01 & 52.7 & 743.2 & 34.5 & 753.1 & 33.6 \\
HD198542& M0III & 828.5 & 35.7 & 15.0 & 3.1 & 5.25E$-$01 & 46.9 & 5.06E$-$01 & 55.2 & 692.1 & 33.3 & 718.0 & 33.6 \\
HD209952& B7IV & 112.7 & 3.8 & 7.5 & 1.1 & 7.86E$-$02 & 57.1 & 7.16E$-$02 & 63.4 & 628.4 & 24.0 & 690.6 & 25.9 \\
HD213310*& M0II+ & 773.3 & 33.9 & 7.5 & 0.5 & 6.66E$-$01 & 66.6 & 6.76E$-$01 & 82.6 & 509.5 & 23.6 & 501.4 & 22.8 \\
HD216131*& G8II & 249.9 & 8.4 & 8.9 & 1.0 & 2.30E$-$01 & 34.6 & 2.21E$-$01 & 40.6 & 476.3 & 21.1 & 495.2 & 20.7 \\
HD217906& M2.5II & 7348.0 & 273.6 & 8.2 & 0.9 & 4.55E+00 & 91.3 & 5.42E+00 & 132.5 & 707.5 & 27.5 & 594.8 & 22.6 \\
\enddata
\tablenotetext{a}{Stars marked with a * were not used for the
final calibration factor as they were rejected as known outliers (\S\ref{sec_results}) or
clipped as being $> 5\sigma$ from the mean (\S\ref{sec_nonlin}).}
\tablenotetext{b}{The measured instrumental flux densities can be
converted to physical units by multiplying by 1.60 Jy MIPS70$^{-1}$.}
\end{deluxetable}


\begin{deluxetable}{llrrrrrrrrrrrr}
\tablewidth{0pt}
\tablecaption{Fine-Scale Calibration Factors \label{tab_ave_nf}}
\tablehead{ & & \multicolumn{2}{c}{Predicted} &
    \multicolumn{2}{c}{Background} & \multicolumn{4}{c}{Average
    Measured\tablenotemark{a}} & \multicolumn{4}{c}{Calibration Factor} \\
    \colhead{Name} & \colhead{SpType} & \colhead{Flux} &
    \colhead{Unc.} &\colhead{SB} & \colhead{Unc} & 
    \colhead{Ap. Flux} & \colhead{S/N} & 
    \colhead{PSF Flux} & \colhead{S/N} & 
    \colhead{Ap.} & \colhead{Unc.} & \colhead{PSF} & \colhead{Unc.}  \\
 & & \multicolumn{2}{c}{[mJy]} & \multicolumn{2}{c}{[MJy sr$^{-1}$]} & 
   \colhead{[MIPS70F]} & &
   \colhead{[MIPS70F]} & & \multicolumn{4}{c}{[MJy sr$^{-1}$ MIPS70F$^{-1}$]}}
\startdata
HD045348& F0II & 3085.0 & 67.1 & 5.6 & 0.5 & 1.66E+00 & 97.6 & 1.56E+00 & 111.0 & 2881.5 & 69.3 & 3066.1 & 72.2 \\
HD048915& A0V & 2900.0 & 354.5 & 14.8 & 1.0 & 1.52E+00 & 100.6 & 1.47E+00 & 118.2 & 2959.1 & 362.9 & 3048.6 & 373.6 \\
HD071129& K3III+ & 5153.0 & 168.7 & 7.5 & 0.6 & 2.90E+00 & 91.3 & 2.68E+00 & 102.0 & 2752.9 & 95.0 & 2980.0 & 101.8 \\
HD080493& K7III & 1686.0 & 60.0 & 10.8 & 1.9 & 9.89E$-$01 & 113.1 & 9.03E$-$01 & 125.0 & 2642.6 & 96.9 & 2892.3 & 105.5 \\
HD082668& K5III & 1424.0 & 204.5 & 25.7 & 0.5 & 9.53E$-$01 & 113.8 & 8.75E$-$01 & 126.5 & 2316.5 & 333.3 & 2522.1 & 362.8 \\
HD100029& M0III & 1124.0 & 38.2 & 4.9 & 0.6 & 6.47E$-$01 & 87.0 & 6.20E$-$01 & 100.8 & 2691.4 & 96.7 & 2809.9 & 99.6 \\
HD108903& M3.5III & 17000.0 & 1766.0 & 18.4 & 0.8 & 1.09E+01 & 150.5 & 9.83E+00 & 163.8 & 2410.5 & 250.9 & 2679.5 & 278.8 \\
HD131873& K4III & 3363.0 & 123.7 & 4.4 & 0.5 & 1.78E+00 & 177.5 & 1.68E+00 & 202.8 & 2925.0 & 108.8 & 3098.2 & 115.0 \\
HD163588& K2III & 354.4 & 9.9 & 4.7 & 0.4 & 2.03E$-$01 & 125.3 & 1.83E$-$01 & 136.4 & 2700.1 & 78.5 & 3002.0 & 86.7 \\
HD217906& M2.5II & 7348.0 & 273.6 & 8.2 & 0.9 & 5.14E+00 & 123.5 & 4.70E+00 & 136.8 & 2215.1 & 84.4 & 2420.0 & 91.8 \\
\enddata
\tablenotetext{a}{The measured instrumental flux densities can be
converted to physical units by multiplying by 1.87 Jy MIPS70$^{-1}$.}
\end{deluxetable}


\begin{deluxetable}{ll}
\tablewidth{0pt}
\tablecaption{Rejected Stars\label{tab_reject}}
\tablehead{\colhead{Name} & \colhead{Reason} }
\startdata
HD 36167 & extended emission \\
HD 35666 & extended emission \\
HD 42701 & extended emission \\
HD 141477 & extended emission \\
HD 166780 & extended emission \\
HD 169916 & extended emission \\
HD 173398 & possible infrared excess \\
HD 173511 & nearby, bright object \\
HD 173976 & extended emission \\
\enddata
\end{deluxetable}

\end{document}